\documentclass[lettersize,journal]{IEEEtran}

\usepackage{amsmath,amsfonts,amssymb,mathtools} 
\usepackage{amsthm} 
\usepackage{xspace} 
\usepackage{graphicx}
\usepackage{subcaption} 
\usepackage{cite}
\usepackage{url}
\usepackage{hyperref}
\usepackage{color}
\usepackage{array}
\usepackage{stfloats}
\usepackage{verbatim}
\usepackage{pifont}
\usepackage{wrapfig}
\usepackage{float}

\usepackage[ruled,vlined,linesnumbered]{algorithm2e}

\usepackage{multirow}
\usepackage{makecell}
\usepackage{booktabs}
\usepackage{tabularx}
\usepackage{threeparttable}

\newtheorem{theorem}{Theorem}

\newcommand{\attcolor}{black}
\newcommand{\workname}{\textit{LOIP}\xspace} 

\allowdisplaybreaks[4] 


\begin{document}

\title{
%
\textit{LOIP}: Collaborative LLM Inference \\ with Offloading-based Pipeline Parallelism \\ on Edge Devices
}

\author{Mingyu Sun, Xiao Zhang, Shen Qu, Yan Li,                                                     Mengbai Xiao, Yanwei Zheng, Yuan Yuan, Dongxiao Yu
\IEEEcompsocitemizethanks{
\IEEEcompsocthanksitem Mingyu Sun, Xiao Zhang, Shen Qu, Yan Li, Mengbai Xiao, and Yanwei Zheng are with the School of Computer Science and Technology, Shandong University, Qingdao 266237, China. Email: sw1nd@mail.sdu.edu.cn, xiaozhang@sdu.edu.cn, sdqushen@mail.sdu.edu.cn, yhyanli@mail.sdu.edu.cn, xiaomb@sdu.edu.cn, and zhengyw@sdu.edu.cn.
\IEEEcompsocthanksitem Yuan Yuan is with the School of Artificial Intelligence, Shandong University, Jinan 250100, China. Email: yyuan@sdu.edu.cn.
\IEEEcompsocthanksitem Dongxiao Yu is with the School of Cryptologic Science and Engineering, Shandong University, Jinan, China, and also with the Shandong Provincial Key Laboratory of Computing-Network Integration, Shandong University, Jinan, China. Email: dxyu@sdu.edu.cn.}
}

\markboth{Journal of \LaTeX\ Class Files,~Vol.~14, No.~8, August~2021}%
{Shell \MakeLowercase{\textit{et al.}}: A Sample Article Using IEEEtran.cls for IEEE Journals}


\maketitle
\begin{abstract}
Deploying large language models (LLMs) on memory-constrained edge devices without model compression remains challenging.
Although model offloading alleviates memory pressure by staging model weights on secondary storage, repeated weight loading introduces substantial pipeline bubbles.
Meanwhile, limited and fluctuating inter-device bandwidth, and continuous KV cache growth further complicate efficient collaborative inference.
To address these challenges, we present \workname{}, a collaborative LLM inference framework that coordinates model offloading with interleaved pipeline parallelism without model compression.
Specifically, \workname{} overlaps weight loading with computation and inter-device communication, using an offloading-based cost model and a fine-grained allocation scheduler to select layer assignments, offloading decisions, and the number of interleaved stages.
A unified memory architecture (UMA)-aware loading mechanism uses customized CUDA kernels to reduce loading overhead, while an online memory adaptation strategy adjusts block-level offloading and KV cache placement as memory pressure and bandwidth change.
We implement \workname{} with more than 2,500 lines of Python and 500 lines of C++/CUDA code and evaluate it on heterogeneous NVIDIA Jetson testbeds using LLMs ranging from 13B to 70B parameters.
Experimental results show that \workname{} achieves up to 8.8$\times$ and 20.3$\times$ speedups over the evaluated baselines under single-input and concurrent-input execution settings, respectively, while leaving the original model parameters unchanged.
\end{abstract}


\section{Introduction}\label{sec:intro}

\IEEEPARstart{R}{ECENTLY}, large language models (LLMs), such as GPT~\cite{radford2018improving}, LLaMA~\cite{dubey2024LLaMA}, and DeepSeek~\cite{liu2024deepseek}, have demonstrated remarkable capabilities in language understanding, reasoning, and decision-making.
Applications such as smart homes and robotics~\cite{king2024sasha,team2025gemini} motivate local inference to support latency-sensitive and privacy-critical tasks while reducing dependence on remote clouds.
However, deploying LLMs on edge devices without model compression remains challenging because their memory and computation requirements substantially exceed the resources available on typical edge platforms.
For example, LLaMA3.3-70B-Instruct requires at least 130~GB of memory, whereas a Jetson Xavier NX provides only 16~GB of unified memory and up to 21~TOPS of AI performance~\cite{nvidia_jetson_xavier_nx}.

\begin{figure}[]
    \centering
    \includegraphics[width=\linewidth]{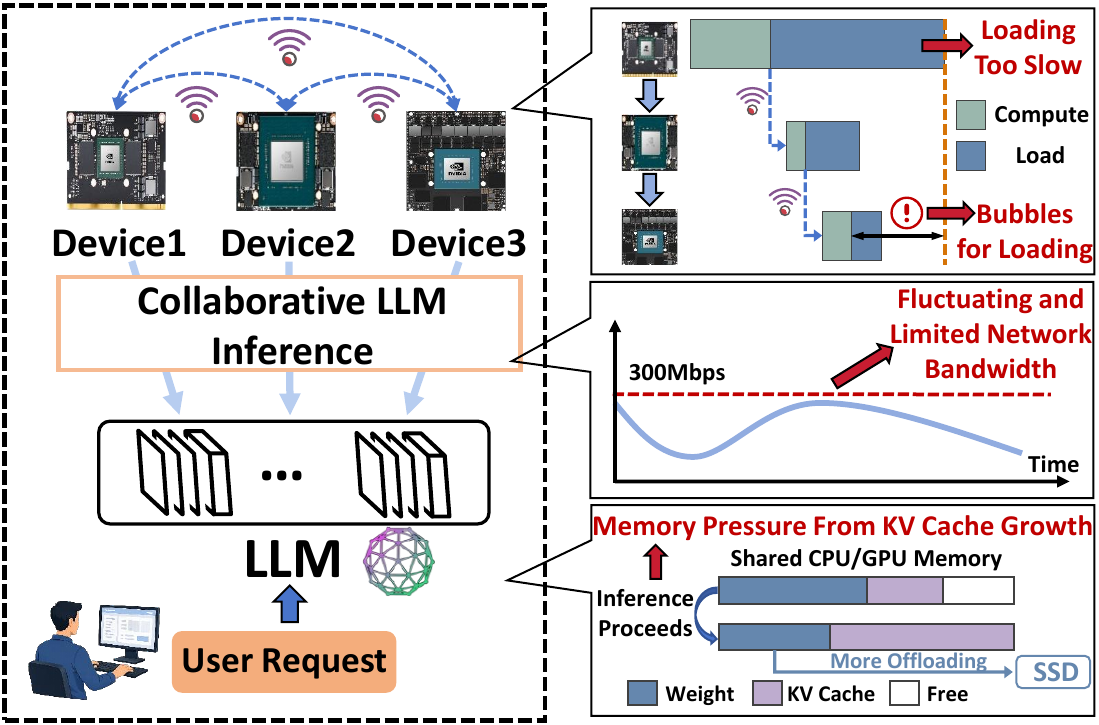}
    \caption{Key challenges in combining model offloading with collaborative LLM inference on resource-constrained edge devices.}
    \label{fig:background}
\end{figure}

A common approach to reducing the deployment cost of LLMs is model compression, including pruning~\cite{li2025unity,ashkboos2024slicegpt,ma2023llm,frantar2023sparsegpt,li2025generalization,wang2023theoretical}, quantization~\cite{hooper2024kvquant,dettmers2023spqr,dettmers2023qlora,frantar2022gptq,lin2024awq}, and distillation~\cite{gu2023minillm,liang2023less,hsieh2023distilling,li2024resource}.
For example, SparseGPT~\cite{frantar2023sparsegpt} removes redundant weights through layer-wise reconstruction, whereas AWQ~\cite{lin2024awq} protects salient weights during low-bit quantization.
Although these techniques reduce the model memory footprint, they modify the original model parameters or numerical representation and may introduce undesirable accuracy degradation in accuracy-sensitive applications~\cite{addo2018credit,li2023medical}.
Model offloading~\cite{hf_accelerate_big_modeling} provides a practical path toward inference while leaving the original model parameters unchanged by storing part of the model weights on SSD and loading them into device memory only when needed, without changing the original parameters.
DeepSpeed Inference~\cite{aminabadi2022deepspeed} organizes model weights across GPU, CPU DRAM, and NVMe SSD tiers, streaming them to GPU on demand during inference.
FlexGen~\cite{sheng2023flexgen} jointly manages weights, activations, and KV cache across GPU, CPU, and disk, and employs a search-based planner to maximize throughput under a given memory budget.
These systems exploit CPU memory as a separate staging tier, whereas unified memory architecture (UMA)-based edge devices have a single physical memory pool shared by the CPU and GPU.
Offloading on UMA-based devices must therefore account for both resident weights and loading buffers within the same limited memory budget.

Collaborative inference provides a complementary approach by aggregating the resources of multiple heterogeneous edge devices.
PipeEdge~\cite{hu2022pipeedge} partitions model layers into pipeline stages according to device computation capabilities.
EdgeShard~\cite{zhang2024edgeshard} jointly optimizes device selection and model partitioning for collaborative LLM inference.
Galaxy~\cite{ye2024galaxy} combines fine-grained workload partitioning with hybrid model parallelism, while TPI-LLM~\cite{li2025tpi} employs tensor parallelism and sliding-window memory management on CPU-oriented edge devices.

However, as illustrated in Fig.~\ref{fig:background}, combining model offloading with collaborative LLM inference introduces three tightly coupled challenges:
(1) \textit{Bubbles for loading}.
When an assigned model slice is unavailable in memory, the corresponding device must wait for the slice to be loaded from SSD before continuing computation.
These loading stalls accumulate across decoding steps and can dominate the overall inference latency.
(2) \textit{Limited and fluctuating inter-device bandwidth}.
Collaborative inference requires intermediate activations and runtime states to be exchanged among devices, and bandwidth variations can disrupt the intended overlap among loading, computation, and communication.
(3) \textit{Continuous KV cache growth}.
As decoding proceeds, the KV cache occupies an increasing fraction of the available memory, leaving less space for retained model weights and potentially forcing additional layers to be offloaded.
This pressure becomes more pronounced when multiple inputs are processed concurrently.
These factors jointly influence layer allocation, offloading, communication, and runtime memory consumption. 
Therefore, a partitioning strategy based solely on computation or memory cannot adequately account for their combined effects on decoding latency.

Thus in this work, we propose \workname{}, an offloading-based collaborative inference framework that combines model offloading with interleaved pipeline parallelism on heterogeneous edge devices without model compression.
The key idea is to distribute each device's assigned layers across interleaved stages, allowing loading for an upcoming stage to overlap with computation and inter-device communication.
Increasing the number of stages creates more opportunities to hide loading latency, but also introduces more inter-device activation transfers.
The design of \workname{} addresses this tradeoff and runtime dynamics through the following three questions:

(1)\textit{How to reduce the loading latency when offloading is combined with collaborative LLM inference?}
Directly inserting offloading into collaborative LLM inference can introduce substantial latency and execution bubbles.
To reduce exposed loading stalls, \workname{} uses an offloading-based cost model and a fine-grained allocation scheduler to select layer assignments, offloading decisions, and the number of interleaved stages based on computation, communication, loading, and memory costs.
Moreover, to reduce the loading overhead itself, \workname{} introduces a runtime-level UMA-aware loading optimization that reuses a mapped buffer and allows customized CUDA kernels to access loaded weights directly.

(2) \textit{How to adapt collaborative inference to fluctuating network bandwidth?}
\workname{} employs a bandwidth-aware KV cache transfer protocol that adjusts the transfer volume according to the available bandwidth.
The protocol temporarily places part of a device's KV cache on a peer with more available memory and retrieves the KV cache before the corresponding stage executes again.
By exploiting communication opportunities during pipeline execution, the protocol balances memory usage and helps delay additional layer offloading on memory-constrained devices.

(3) \textit{How to accommodate continuous KV cache growth during inference?}
\workname{} employs an online memory-aware planner that uses device-specific token thresholds to track memory pressure and progressively adjusts block-level offloading when these thresholds are reached.
This mechanism releases memory occupied by retained weights to accommodate the growing KV cache, while accounting for the additional weight-loading overhead.
Together with bandwidth-aware KV cache transfers, the planner adapts weight residency to memory growth under both single-input and concurrent-input execution.

The main contributions of this work are summarized as follows:

\begin{itemize}
\setlength{\itemsep}{0pt}
\setlength{\parsep}{0pt}
\setlength{\parskip}{0pt}

\item We propose \workname{}, an offloading-based interleaved pipeline parallelism framework for collaborative LLM inference on heterogeneous, memory-constrained edge devices without model compression.
At its core, \workname{} uses interleaved pipeline parallelism to overlap weight loading with computation and inter-device communication, thereby reducing exposed loading stalls.

\item The core design of \workname{} addresses three key challenges: \textit{reducing latency introduced by offloading}, \textit{handling network bandwidth variations} and \textit{accommodating continuous KV cache growth}.
Specifically, \workname{} develops an offloading-based cost model to characterize inference latency and memory requirements, and uses a fine-grained allocation scheduler to determine layer assignments, offloading decisions, and the number of interleaved stages.
A UMA-aware loading optimization with customized CUDA kernels reduces runtime loading overhead, while an online memory adaptation strategy coordinates block-level offloading and KV cache placement as memory pressure and bandwidth change.

\item We implement \workname{} on real-world heterogeneous NVIDIA Jetson testbeds with 2500+ lines of Python and 500+ lines of C++/CUDA code.
Extensive experiments demonstrate that \workname{} achieves 8.8$\times$ and 20.3$\times$ speedups over the SOTA baselines under sporadic and bursty request patterns respectively, while leaving the original model parameters unchanged.
\end{itemize}

\IEEEpubidadjcol

\section{Related Work}
\subsection{Memory Optimization for Efficient LLM Inference}\label{light_weight}
To facilitate efficient inference of LLMs on resource-constrained devices, substantial efforts have been devoted to memory-efficient LLM inference techniques, including model compression and offloading-based inference.
Model compression techniques, such as quantization~\cite{hooper2024kvquant,dettmers2023spqr,dettmers2023qlora,frantar2022gptq,lin2024awq}, distillation~\cite{gu2023minillm,liang2023less,hsieh2023distilling,li2024resource}, and pruning~\cite{li2025unity,ashkboos2024slicegpt,ma2023llm,frantar2023sparsegpt,li2025generalization,wang2023theoretical}, aim to reduce the computational and memory footprint of LLMs during inference.
Specifically, GPTQ~\cite{frantar2022gptq} minimizes quantization error via layer-wise greedy optimization.
AWQ~\cite{lin2024awq} preserves model performance under low-bit quantization by applying activation-aware weight scaling.
MiniLLM~\cite{gu2023minillm} uses reverse KL distillation with teacher-guided sampling to reduce model size.
SparseGPT~\cite{frantar2023sparsegpt} proposes a post-training pruning method.
KVQuant~\cite{hooper2024kvquant} quantizes the KV cache during inference, whereas QLoRA~\cite{dettmers2023qlora} enables memory-efficient fine-tuning of quantized LLMs.
Although these methods reduce memory consumption, they inevitably compromise model accuracy and require intrusive modifications to the original model.
For offloading-based inference systems, DeepSpeed-Inference~\cite{aminabadi2022deepspeed} supports inference under limited GPU memory by utilizing CPU memory and NVMe storage.
FlexGen~\cite{sheng2023flexgen} formulates LLM inference as a tensor placement and scheduling problem across GPU, CPU, and disk.
The use of CPU memory as a separate staging tier in these offloading systems differs from UMA-based edge execution, where CPU and GPU allocations share the same physical memory budget.
vLLM~\cite{kwon2023efficient} introduces PagedAttention to efficiently manage KV cache memory, but mainly targets cloud-based scenarios.
However, these systems are designed for cloud platforms with abundant CPU memory, making them unsuitable for resource-constrained UMA-based edge devices.

\begin{table}[t]
\caption{Feature comparison of collaborative LLM inference systems.}
\label{tab:related_work_table}
\centering
\renewcommand{\arraystretch}{1.15}
\setlength{\tabcolsep}{4pt}
\resizebox{0.95\columnwidth}{!}{%
\begin{tabular}{lccccc}
\toprule
\textbf{Method} & \textbf{Type} & \textbf{Backends} & 
\makecell{\textbf{Model Memory}\\\textbf{Footprint}}&
\makecell{\textbf{Online}\\\textbf{Adaptation}} & \makecell{\textbf{UMA}\\\textbf{Optimization}} \\
\midrule
PipeEdge  & PP    & CPU     & ---   & \ding{55} & \ding{55} \\
EdgeShard & PP    & GPU     & ---   & \ding{55} & \ding{55} \\
Jupiter   & PP    & GPU     & $\uparrow$ & \ding{55} & \ding{55} \\
TPI-LLM   & TP    & CPU     & $\downarrow$  & \ding{52} & \ding{55} \\
Galaxy    & TP+SP & CPU/GPU & ---   & \ding{55} & \ding{55} \\
\midrule
\workname{} & PP  & GPU     & $\downarrow$  & \ding{52} & \ding{52} \\
\bottomrule
\end{tabular}}
\end{table}

\subsection{Collaborative Inference on Edge} \label{edge_infer}
An alternative and complementary approach is to unify heterogeneous edge devices to enable collaborative inference of LLMs, thereby achieving more efficient on-device inference.
Early studies mainly focus on distributing conventional DNN workloads.
Band~\cite{jeong2022band} coordinates multiple DNNs across heterogeneous mobile processors, CoEdge~\cite{zeng2020coedge} adapts DNN partitioning to computation and network conditions, and Voltage~\cite{hu2024edge} extends automatic partitioning to Transformer models.
For larger models, PipeEdge~\cite{hu2022pipeedge} maps pipeline stages according to heterogeneous device capabilities, while EdgeShard~\cite{zhang2024edgeshard} jointly optimizes device selection and layer partitioning for collaborative LLM inference.
Galaxy~\cite{ye2024galaxy} combines fine-grained workload partitioning with hybrid model parallelism, and Jupiter~\cite{ye2025jupiter} further improves fine-grained parallel execution and KV cache sharing.
TPI-LLM~\cite{li2025tpi} instead targets CPU-oriented edge platforms using tensor parallelism and sliding-window memory management.

However, these systems differ from \workname{} in their assumptions about model residency, hardware backends, and runtime adaptation.
Table~\ref{tab:related_work_table} summarizes their features.
Jupiter~\cite{ye2025jupiter} incorporates speculative decoding through auxiliary self-drafting heads, which introduce additional resident memory overhead.
Although its partition planner accounts for device memory budgets, it assumes that the model partition assigned to each device remains resident in memory.
TPI-LLM~\cite{li2025tpi} employs sliding-window weight offloading and runtime memory scheduling, but is designed primarily for CPU-oriented edge devices.
Moreover, it does not coordinate weight residency with continuous KV cache growth or adapt its communication policy to runtime bandwidth variations.
Existing work therefore leaves open how to jointly schedule layer placement, repeated weight loading, inter-device communication, and continuous KV cache growth across heterogeneous, memory-constrained edge devices, which is the gap addressed by \workname{}.

\begin{figure*}[t]
    \centering
    \subfloat[Layer allocation and computing timeline of conventional pipeline parallelism.]{%
      \includegraphics[width=0.9\linewidth]{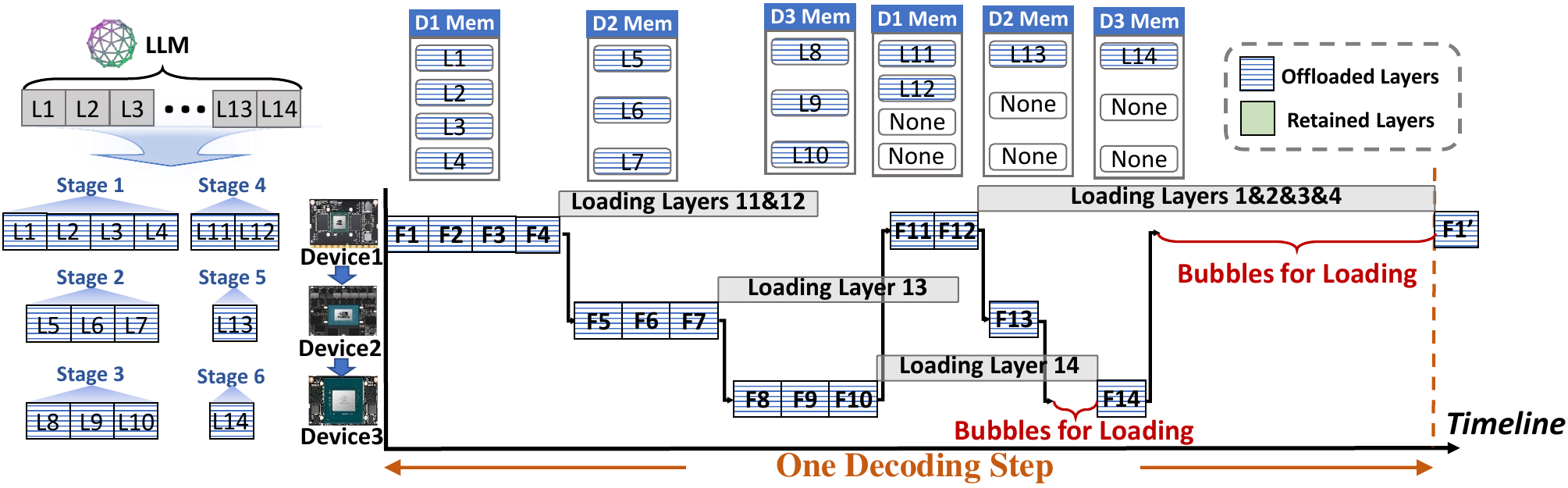}
      \label{fig:trad}
    }\vspace{0.5em}

    \subfloat[Layer allocation and computing timeline of interleaved pipeline parallelism.]{%
      \includegraphics[width=0.9\linewidth]{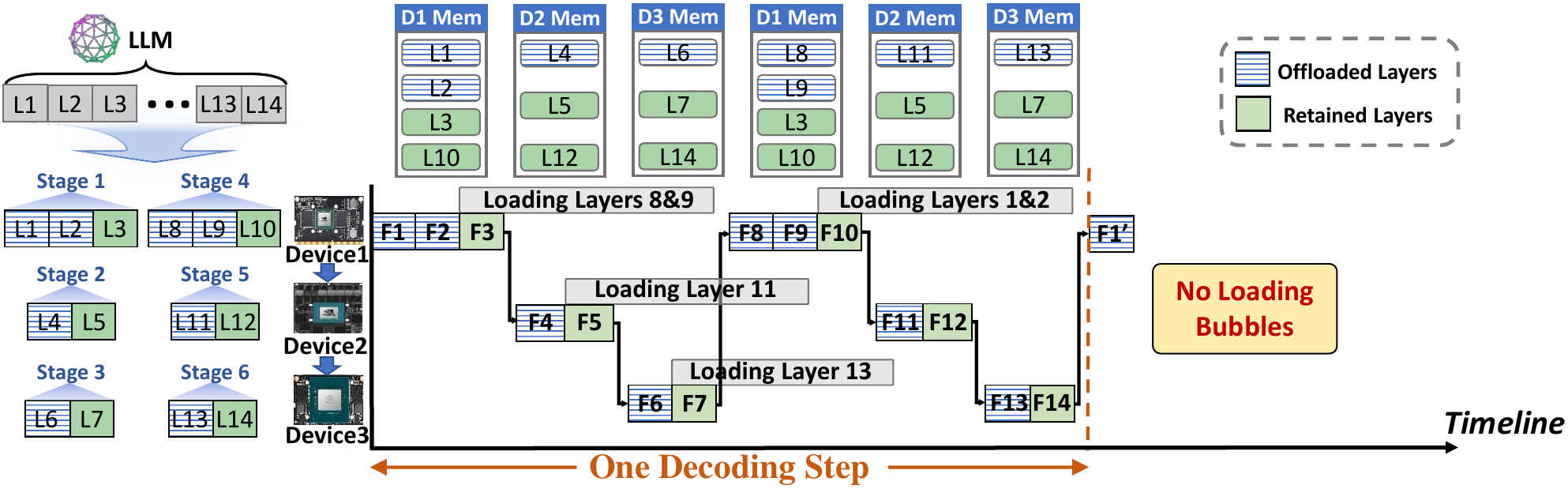}
      \label{fig:loip}
    }
    \caption{Comparison between the conventional pipeline parallelism and interleaved pipeline parallelism. FN represents the forward pass of Layer N and FN' represents the forward pass of Layer N at the next decoding step.}
    \label{fig:inter_run}
\end{figure*}
 
\section{System Design}\label{sec:framework_design}

\subsection{Overview of \workname{}}\label{sub:overview}
Given a set of heterogeneous edge devices $\mathcal{D}$ and the LLM decoder layers $\mathcal{L}$, \workname{} aims to enable collaborative LLM inference on resource-constrained edge devices while minimizing inference latency.
For the device $i$, its available memory capacity is denoted as $Mem_i$, and the layers assigned to it are denoted as $\mathcal{L}_i$.
Due to the limited unified memory consumed by model weights and KV cache, keeping all assigned layers retained in memory is often infeasible.
\workname{} therefore keeps a subset of these layers resident and stores the remaining layers $\Tilde{\mathcal{L}}_i \subseteq \mathcal{L}_i$ on SSD for loading when required.

As shown in Fig.~\ref{fig:trad}, since device memory cannot accommodate all model layers, a straightforward conventional pipeline first assigns each device a subset of layers matching its computing capability and memory budget.
After computing these layers, all model weights in device memory must be offloaded, and the remaining layers are then loaded before subsequent computation can proceed.
However, this approach introduces significant pipeline bubbles due to loading stalls, which substantially increases the inference latency.

To address this issue, \workname{} adopts an offloading-based interleaved pipeline parallelism, as shown in Fig.~\ref{fig:loip}.
In the interleaved pipeline, retained and offloaded layers are distributed as evenly as possible across the stages of each device.
Once a device finishes executing the offloaded layers of its current stage, it can reuse the loading buffer to load weights for its next stage.
This loading can overlap with computation on the current stage's retained layers, execution on other devices, and inter-device communication, thereby reducing exposed loading latency.
Fig.~\ref{fig:loip} shows an interleaved pipeline where each device contains two stages.
For Device 1, Layer 3 and Layer 10 are kept in device memory and do not need to be offloaded during pipeline execution.
They occupy the same relative position in their respective stages, so that the loading time of dynamically loaded layers is uniformly interleaved with computation and communication across stages.

\begin{figure*}[t]
    \centering
    \includegraphics[width=0.95\linewidth]{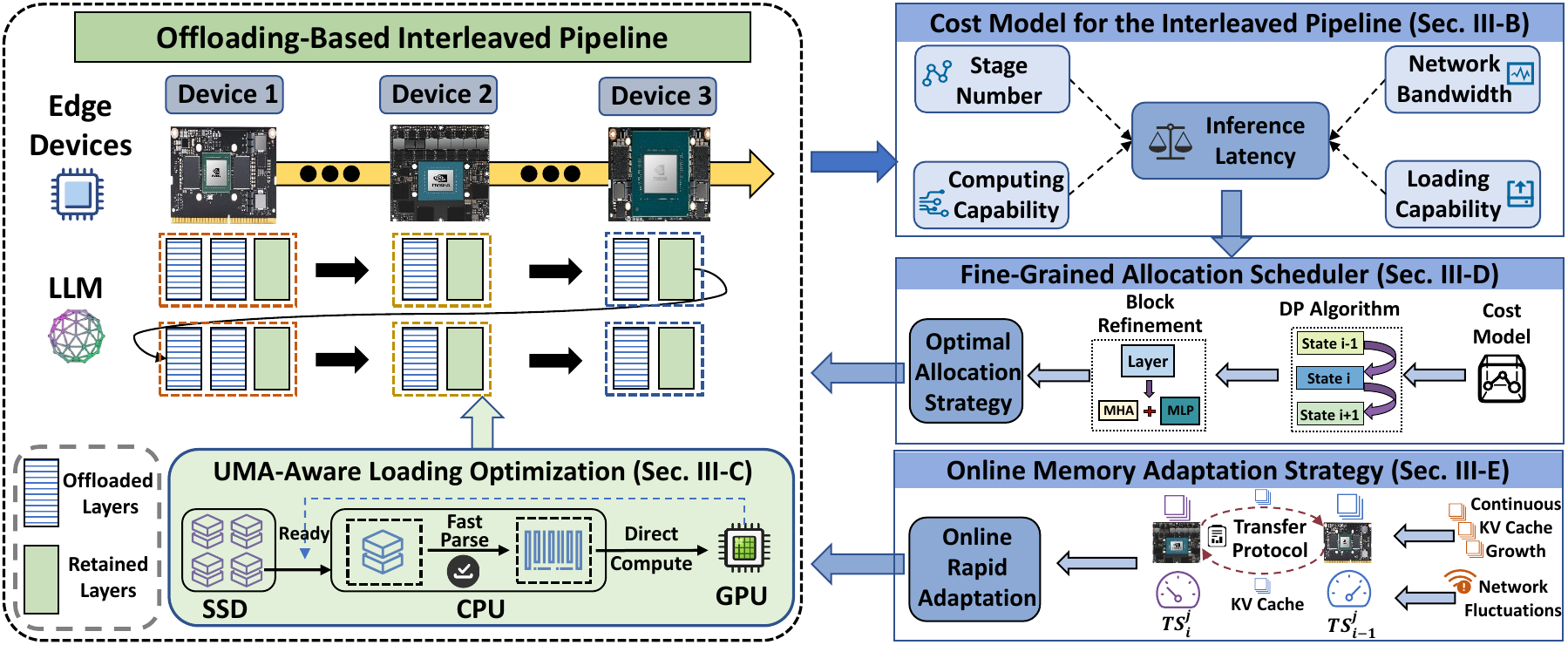}
    \caption{Overview of the \workname{} framework. At its core, \workname{} adopts interleaved pipeline parallelism to support collaborative inference across resource-constrained edge devices, comprising cost model construction, UMA-aware loading optimization, fine-grained allocation scheduler, and online memory adaptation strategy.}
    \label{fig:overall_framework}
\end{figure*}

Based on this interleaved pipeline parallelism, \workname{} comprises four components, as shown in Fig.~\ref{fig:overall_framework}.
First, an offloading-based cost model (Sec.~\ref{cost}) characterizes computing, communicating, loading, and memory overhead, and computes the overlap window available on each device.
Then, based on these costs and overlap windows, the fine-grained allocation scheduler (Sec.~\ref{scheduler}) selects the number of stages per device, assigns layers to devices, and determines which layers or blocks are loaded on demand.
Next, during execution, the UMA-aware loading optimization (Sec.~\ref{uma}) uses a reusable mapped loading buffer and customized CUDA kernels to reduce the overhead of repeatedly loading model slices.
Finally, the online memory adaptation strategy (Sec.~\ref{online}) adjusts block-level offloading and KV cache placement as the KV cache grows and network bandwidth changes.
Together, these components allow \workname{} to support LLM inference across resource-constrained edge devices while leaving the original model parameters unchanged.

\subsection{Cost Model for the Interleaved Pipeline Parallelism}\label{cost}
To quantify the inference latency of the proposed interleaved pipeline parallelism, we develop a heterogeneous offloading-based cost model. 
We consider a heterogeneous, memory-constrained edge system comprising \(|\mathcal{D}|\) devices.
For each device \(i \in \mathcal{D}\), \(Mem_i\) denotes the memory budget available for decoder weights and KV cache after accounting for other memory usage, including the operating system, runtime, and activations.
The devices communicate over inter-device links with effective bandwidth $bw_{net}$.
Let $\mathcal{L}_i$ denote all layers assigned to the device $i$, and $\Tilde{\mathcal{L}}_i \subseteq \mathcal{L}_i$ denote its offloaded layers.
The assigned layer sets form a disjoint partition of $\mathcal{L}$, and execution preserves the original layer order.
All devices use the same number of stages, denoted by $\#Stage$.
For stage $s$ of the device $i$, let $\mathcal{L}_{i,s}$ be its layer set and $\Tilde{\mathcal{L}}_{i,s}=\Tilde{\mathcal{L}}_i\cap\mathcal{L}_{i,s}$ its offloaded subset.
The stage sets form a disjoint partition of $\mathcal{L}_i$ and each stage contains at least one layer, but its offloaded subset may be empty.
Accordingly, the optimization objective for the interleaved pipeline parallelism is formulated as follows:
\begin{align}
\label{eq:cost_model}
&\min \quad T_{total} 
   = \underbrace{\sum_{i \in \mathcal{D}} \textit{comp}(\mathcal{L}_i)}_{T_{comp}}
    + \underbrace{\#Stage \cdot \frac{|\mathcal{D}| \cdot h_{\mathrm{size}}}{bw_{net}}}_{T_{comm}} \\
   &\left.
      \begin{aligned}
+\#Stage &\cdot \mathop{\max}\limits_{i \in \mathcal{D}} \Big\{ \max \Big\{ 
            \frac{\textit{load}(\Tilde{\mathcal{L}}_i)}{\#Stage} - \frac{\textit{comp}(\mathcal{L}_i - \Tilde{\mathcal{L}}_i) }{\#Stage} \nonumber\\
        &\quad- \frac{\sum_{i' \in \mathcal{D}-\{i\}} \textit{comp}(\mathcal{L}_{i'})}{\#Stage}
          - \frac{|\mathcal{D}| \cdot h_{\mathrm{size}}}{bw_{net}},\, 0 \Big\} \Big\}
      \end{aligned}
      \right\}
      T_{uncover} ,\nonumber\\[1.2ex]
&\text{s.t.} \quad
   \Tilde{\mathcal{L}}_i \subseteq \mathcal{L}_i, \quad \forall i \in \mathcal{D}, \nonumber\\
        &\qquad 0 \leq\; (|\mathcal{L}_i|-|\Tilde{\mathcal{L}}_i|)\cdot l_{\mathrm{size}}+ \max_{1\leq s\leq\#Stage}|\Tilde{\mathcal{L}}_{i,s}|\cdot l_{\mathrm{size}} \nonumber\\
        &\qquad \qquad + M_i^{\mathrm{KV}}(n) 
        \leq Mem_i,\quad \forall i \in \mathcal{D}, \nonumber\\
   &\qquad 2\leq\#Stage\leq\min_{i\in\mathcal{D}}|\mathcal{L}_i|,
   \quad \#Stage\in\mathbb{Z}. \nonumber
\end{align}
Here, $T_{total}$ is the modeled latency of one decoding step.
$T_{comp}$, $T_{comm}$, and $T_{uncover}$ indicate the computing time, communication time, and uncovered loading time, respectively.

For $T_{comp}$, $\textit{comp}(\mathcal{L}_i)$ denotes the computing time of the assigned layers $\mathcal{L}_i$ on the device $i$. 
For $T_{comm}$, $h_{\mathrm{size}}$ denotes the size of the hidden states transferred at each stage boundary during a single decoding step.
One round executes one stage on each device and includes $|\mathcal{D}|$ hidden state transfers, including the transfer from the last device back to the first.
Therefore, the total communication time overhead for a single decoding step is $\#Stage \cdot \frac{|\mathcal{D}| \cdot h_{\mathrm{size}}}{bw_{net}}$.

Considering $T_{uncover}$, the loading time of the device $i$ can be overlapped with the computing time of its remaining layers, the computing time of other devices, and the communication time. Specifically, $\textit{load}(\Tilde{\mathcal{L}}_i)$ denotes the time required for the device $i$ to load layers $\Tilde{\mathcal{L}}_i$, and
$\textit{comp}(\mathcal{L}_i - \Tilde{\mathcal{L}}_i)$ denotes the computing time of the layers on the device $i$ that do not need to be offloaded. 
Additionally, $\sum_{i' \in \mathcal{D} - \{i\}} \textit{comp}(\mathcal{L}_{i'})$ represents the computing time of all devices in the pipeline except device $i$.
Moreover, the interleaved pipeline parallelism enables the loading time across edge devices to be overlapped to eliminate bubbles for loading.
Consequently, the overall system loading time is given by the maximum loading time among all devices.

In the interleaved pipeline, the offloaded layers $\Tilde{\mathcal{L}}_i$ of the device $i$ are evenly distributed across $\#Stage$.
When the layer counts are not divisible by $\#Stage$, the actual stages may differ by one layer.
During execution, each device only keeps its retained layers and the offloaded layers required by the current stage in memory.
Thus, the model-weight memory footprint of the device $i$ consists of the retained layers $(|\mathcal{L}_i|-|\Tilde{\mathcal{L}}_i|)\cdot l_{\mathrm{size}}$ and the maximum offloaded layers in a stage $\max_{1\leq s\leq\#Stage}|\Tilde{\mathcal{L}}_{i,s}|\cdot l_{\mathrm{size}}$.
For the KV cache, $n$ denotes the cached sequence length, including both input and generated tokens, and $M_i^{\mathrm{KV}}(n)$ denotes the peak resident KV cache memory on the device $i$ during the decoding step.
This peak includes the device's own KV cache and the KV cache temporarily stored for other devices.
If the KV cache of the device $i$ needs to be temporarily stored on another device, this value will be less than the memory occupied by its own KV cache.
The retained weights, loading buffer, and peak resident KV cache must together fit within the available budget $Mem_i$.
Finally, every stage must contain at least one layer, giving $\#Stage\leq\min_{i\in\mathcal{D}}|\mathcal{L}_i|$.
And $\#Stage\geq2$ specifies the interleaved configurations considered by the model.
The frequently used notations are listed in Tab.~\ref{table:notation}.

\subsection{UMA-Aware Loading Optimization}\label{uma}

\begin{figure}[t]
  \centering
  \begin{subfigure}[t]{\linewidth}
    \centering
    \includegraphics[width=\linewidth]{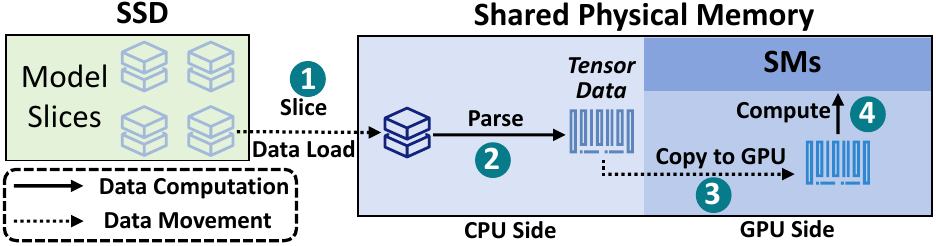}
    \caption{Conventional data path for model slice loading and computing.}
    \label{fig:load_opti1}
  \end{subfigure}
  \vspace{0.5em}
  \begin{subfigure}[t]{\linewidth}
    \centering
    \includegraphics[width=\linewidth]{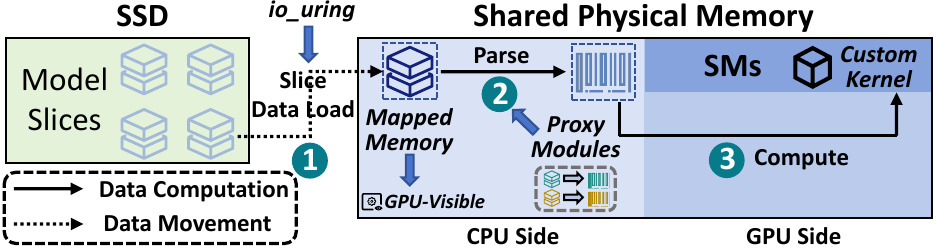}
    \caption{Optimized data path for model slice loading and computing in \workname{}.}
    \label{fig:load_opti2}
  \end{subfigure}
  \caption{Comparison between the conventional and \workname{}-optimized model slice loading and computing.}
\end{figure}

For resource-constrained edge devices, offloaded model layers must be repeatedly loaded from SSD across decoding steps. 
The conventional loading path introduces substantial additional latency in this scenario, as shown in Fig.~\ref{fig:load_opti1}.
The conventional loading path first reads slice data from SSD storage (step~\ding{172}), parses the slice data into computing-ready tensor data in CPU memory (step~\ding{173}), explicitly copies these tensors to GPU memory (step~\ding{174}), and finally executes computation on GPU streaming multiprocessors (SMs) (step~\ding{175}).
When invoked on every decoding step, this path exposes substantial loading latency.
To further mitigate the repeated loading of known layers at every decoding step, \workname{} replaces it with a single-slot direct-access loading runtime.

Fig.~\ref{fig:load_opti2} shows the optimized loading path in \workname{}.
At runtime initialization, \workname{} allocates one pinned host-memory buffer, maps it into the GPU address space, and sizes it to accommodate the largest registered model slice (step~\ding{172}).
The loading backend uses io\_uring~\cite{io_uring} to read each slice's weight bytes into this buffer.
The buffer remains allocated across reloads, while its contents are replaced according to the execution schedule.
This gives file loading and GPU computation access to the same weight storage and keeps the device-visible base address fixed for subsequent executions.

The fixed buffer address also enables \workname{} to reuse the execution structures associated with each slice (step~\ding{173}).
When a slice is first activated, \workname builds and caches lightweight proxy modules for the slice.
These proxies retain the original module configuration and tensor dimensions, but their linear operators hold addressing metadata instead of separate weight tensors.
When the same slice is loaded again, its weights occupy the same offsets in the buffer, allowing the cached proxies to be reused without reconstructing modules or materializing their weights as separate tensors.
Proxy construction is therefore performed once per slice within a runtime configuration, rather than once per decoding step.

To use these weights for GPU computation, \workname{} implements a custom CUDA operator for the proxies' linear projections (step~\ding{174}).
The operator obtains each weight address from the mapped buffer's device-visible base address and the tensor's cached byte offset, then supplies this address as the weight operand to a cuBLAS matrix multiplication on the current CUDA stream.
Thus, the loaded weights can serve as GPU computation inputs without first being copied into a separate GPU weight allocation.
After submitting the computation that consumes the current slice, the runtime records a CUDA event~\cite{cuda_event} on the computation stream.
A background loading thread waits for this event to complete before releasing the current slice's handle and loading the next slice into the same buffer.
Loading can then overlap with subsequent resident-layer computation and execution on other devices, neither of which requires the buffer's current contents.
Before the next slice is consumed, the execution thread waits for loading to finish and activates its cached proxies.
This coordination supports overlap between I/O and GPU execution using a single weight buffer, avoiding a second buffer that would consume shared memory otherwise available for resident weights and KV cache.

\begin{table}[t]
 \caption{\textcolor{\attcolor}{Frequently used notations and corresponding descriptions.}}
 \label{table:notation}
 \centering
	\begin{tabular}[t]{p{1.3cm} p{6.5cm}}
		\hline
		\textcolor{\attcolor}{Notations} & \textcolor{\attcolor}{Descriptions} \\ 
		\hline
            \textcolor{\attcolor}{$\mathcal{L}$} & \textcolor{\attcolor}{Decoder layers of the LLM}\\
            \textcolor{\attcolor}{$\mathcal{D}$} & \textcolor{\attcolor}{Set of heterogeneous edge devices}\\
            \textcolor{\attcolor}{$bw_{net}$} & \textcolor{\attcolor}{Inter-device network bandwidth}\\
            \textcolor{\attcolor}{$Mem_i$} & \textcolor{\attcolor}{Memory budget for decoder weights and KV cache on device $i$}\\
            \textcolor{\attcolor}{$\#Stage$} & \textcolor{\attcolor}{Number of stages per device}\\
            \textcolor{\attcolor}{$\mathcal{L}_i$} & \textcolor{\attcolor}{Layers assigned to device $i$}\\
            \textcolor{\attcolor}{$\mathcal{L}_{i,s}$} & \textcolor{\attcolor}{Layers assigned to stage $s$ of device $i$}\\
            \textcolor{\attcolor}{$\Tilde{\mathcal{L}}_i$} & \textcolor{\attcolor}{Layers assigned to device $i$ for offloading}\\
            \textcolor{\attcolor}{$\Tilde{\mathcal{L}}_{i,s}$} & \textcolor{\attcolor}{Offloaded layers of stage $s$ on device $i$}\\
            \textcolor{\attcolor}{$l_{\mathrm{size}}$} & \textcolor{\attcolor}{Weight size of one decoder layer}\\
            \textcolor{\attcolor}{$h_{\mathrm{size}}$} & \textcolor{\attcolor}{Size of hidden states transferred at each stage boundary}\\
            \textcolor{\attcolor}{$\textit{comp}(\cdot)$} & \textcolor{\attcolor}{Profiled computing time for a layer or block set}\\
            \textcolor{\attcolor}{$\textit{load}(\cdot)$} & \textcolor{\attcolor}{Profiled loading time for a layer or block set}\\
            \textcolor{\attcolor}{$q_i(k)$} & \textcolor{\attcolor}{Minimum offloaded layers when device $i$ is assigned $k$ layers}\\
            \textcolor{\attcolor}{$b_i$} & \textcolor{\attcolor}{Offloaded-layer computing and loading cost on device $i$}\\
            \textcolor{\attcolor}{$b_i(k)$} & \textcolor{\attcolor}{Computing and loading cost of the offloaded layers when device $i$ is assigned $k$ layers}\\
            \textcolor{\attcolor}{$\mathcal{B}_s$} & \textcolor{\attcolor}{Candidate $b_i$ values for stage count $s$}\\
            \textcolor{\attcolor}{$F_{\tau}(l,i)$} & \textcolor{\attcolor}{Minimum compute time for assigning $l$ layers to the first $i$ devices subject to bound $\tau$}\\
            \textcolor{\attcolor}{$P_{\tau}(l,i)$} & \textcolor{\attcolor}{Layer count selected for device $i$ at DP state $(l,i)$ for backtracking}\\
            \textcolor{\attcolor}{$M_i^W$} & \textcolor{\attcolor}{Peak model weight footprint on device $i$ under the current plan}\\
            \textcolor{\attcolor}{$M_i^{\mathrm{KV}}(n)$} & \textcolor{\attcolor}{Peak resident KV cache memory on device $i$}\\
            \textcolor{\attcolor}{$TS^j_i$} & \textcolor{\attcolor}{Token length threshold for the $j$-th offloading trigger on device $i$}\\
            \textcolor{\attcolor}{$V_i^{trans}$} & \textcolor{\attcolor}{Token budget for KV cache transfer across stages of device $i$}\\
            \textcolor{\attcolor}{$p_A,p_M$} & \textcolor{\attcolor}{Memory proportion of the MHA and MLP blocks in one layer}\\
            \textcolor{\attcolor}{$\alpha,\beta$} & \textcolor{\attcolor}{Numbers of additional MHA and MLP blocks offloaded per stage}\\
		\hline
	\end{tabular}
\end{table}

\subsection{Fine-Grained Allocation Scheduler}\label{scheduler}
A straightforward approach to model slice allocation is to exhaustively enumerate all possible assignment combinations and select the optimal one. 
However, for LLMs with many layers deployed across multiple devices, the search space of feasible allocations becomes prohibitively large, resulting in excessive model partitioning time.

To address this challenge, we propose a fine-grained allocation scheduler that efficiently searches the modeled allocation space.
The scheduler jointly selects the stage count, layer assignment, and weights to offload (Alg.~\ref{algo:allocation}).
For each stage count, dynamic programming finds an optimal whole-layer allocation.
A lightweight greedy pass then uses remaining memory to keep selected MHA or MLP blocks resident and reduce repeated loading.

\workname uses the cost model in Sec.~\ref{cost} and assumes equal-cost decoder layers.
For any decoder layer $\ell\in\mathcal{L}$, $\textit{comp}(\{\ell\})$ and $\textit{load}(\{\ell\})$ denote its profiled computation and loading times on the currently considered device $i$.
Accordingly, the computation and loading costs of $k$ layers are $k\cdot\textit{comp}(\{\ell\})$ and $k\cdot\textit{load}(\{\ell\})$, respectively.
The scheduler assumes no KV cache occupancy, and subsequent KV cache growth is handled by the online memory adaptation strategy in Sec.~\ref{online}.
Since each stage must be nonempty, the scheduler enumerates $s=2,\ldots,\lfloor|\mathcal{L}|/|\mathcal{D}|\rfloor$ and lets the DP distribute all layers across devices.

For fixed $s$, assigning $k\geq s$ layers to the device $i$ and offloading $q$ of them requires
\begin{equation}
\label{eq:unit_memory}
 \left(k-q+\left\lceil\frac{q}{s}\right\rceil\right) \cdot
 l_{\mathrm{size}}\leq Mem_i.
\end{equation}
$k-q$ counts resident layers, while $\lceil q/s\rceil$ is the largest loading unit when offloaded layers are balanced across stages.
Let $q_i(k)$ be the smallest integer $q\in[0,k]$ satisfying this constraint.
These counts are precomputed for the current $s$. For example, memory for four layers can support six layers in two stages by keeping two layers resident and reloading two per stage. Thus, $q_i(6)=4$ counts all reloaded layers, including those whose memory is reused as the shared buffer.

\begin{algorithm}[t]
  \small
  \caption{Layer Allocation with Block-Level Refinement}
  \label{algo:allocation}
  \SetKwInput{Input}{Input}
  \SetKwInput{Output}{Output}
  \Input{Layers $\mathcal{L}$, devices $\mathcal{D}$, memory budgets, layer/block sizes, and computation, loading, and communication costs}
  \Output{Allocation $\mathcal{A}$}
  $\mathcal{A}\gets\emptyset$\\
  \For{$s=2$ \KwTo $\lfloor|\mathcal{L}|/|\mathcal{D}|\rfloor$}{
    Precompute feasible $q_i(k)$ and $b_i(k)$ for $s\leq k\leq |\mathcal{L}|-(|\mathcal{D}|-1)s$, and $T_{comm}$ for $s$\\
    $\mathcal{B}_s\gets$ distinct precomputed values of $b_i(k)$\\
    $T_s\gets\infty$\\
    \ForEach{$\tau\in\mathcal{B}_s$}{
      Set $F_{\tau}(0,0)=0$, all other states to $\infty$, and all entries of $P_{\tau}$ to 0\\
      \For{$i=1$ \KwTo $|\mathcal{D}|$}{
        \For{$l=is$ \KwTo $|\mathcal{L}|-(|\mathcal{D}|-i)s$}{
          \For{$k=s$ \KwTo $l-(i-1)s$ with feasible memory and $b_i(k)\leq\tau$}{
            $v\gets F_{\tau}(l-k,i-1)+k\cdot\textit{comp}(\{\ell\})$\\
            \If{$v<\infty$ and either $v<F_{\tau}(l,i)$ or [$v=F_{\tau}(l,i)$ and $k>P_{\tau}(l,i)$]}{
              Set $F_{\tau}(l,i)\gets v$ and $P_{\tau}(l,i)\gets k$\\
            }
          }
        }
      }
      $T_{\tau}\gets\max\{F_{\tau}(|\mathcal{L}|,|\mathcal{D}|)+T_{comm},\tau\}$\\
      \If{$F_{\tau}(|\mathcal{L}|,|\mathcal{D}|)<\infty$ and $T_{\tau}<T_s$}{
        Backtrack through $P_{\tau}$ to construct $\mathcal{A}_s$, and set $T_s\gets T_{\tau}$\\
      }
    }
    \If{$T_s<\infty$}{
      Construct balanced stages of $\mathcal{A}_s$, preserving layer order\\
      $Q\gets$ max-heap of devices keyed by $b_i$\\
      \While{$Q\neq\emptyset$}{
        $i\gets Q.\operatorname{pop\_max}()$\\
        \If{an MLP block of a fully offloaded layer fits}{
          Keep the MLP block resident and reload only its MHA block
        }
        \ElseIf{an MHA block of a fully offloaded layer fits}{
          Keep the MHA block resident and reload only its MLP block
        }
        \Else{\textbf{break}}
        Update resident weights, largest unit, and $b_i$\\
        Insert device $i$ back to $Q$\\
      }
      \If{$\mathcal{A}=\emptyset$ or $\mathcal{A}_s$ has lower $T_{total}$}{ $\mathcal{A}\gets\mathcal{A}_s$ }
    }
  }
  \Return{$\mathcal{A}$}
\end{algorithm}

\textbf{Matching the cost model}.
Let $b_i$ denote the loading time plus the computation time of the device $i$'s offloaded weights over one decoding step. Rearranging the uncovered-loading term in Eq.~\ref{eq:cost_model} gives
\begin{equation}
\label{eq:equivalent_latency}
 T_{total}=\max\left\{T_{comp}+T_{comm},\ \max_i b_i\right\}.
\end{equation}
For a whole-layer assignment, $b_i(k)=q_i(k)\cdot\bigl[\textit{comp}(\{\ell\})+\textit{load}(\{\ell\})\bigr]$, which contains only offloaded-layer costs after the computation of resident layers and other devices cancels against the corresponding terms in the loading-overlap window.
Computation of all layers remains included in $T_{comp}$.
Using the smallest feasible $q_i(k)$ minimizes this local cost without changing computation.
The communication term $T_{comm}$ is fixed for a given $s$.
This formulation accounts for overlap through the original cost model without requiring overlap windows for incomplete assignments.

\textbf{Dynamic programming.}
As shown in Alg.~\ref{algo:allocation}, for a fixed stage count $s$, each of the other $|\mathcal{D}|-1$ devices must receive at least $s$ layers.
Thus, a device can receive at most $|\mathcal{L}|-(|\mathcal{D}|-1)s$ layers, and $\mathcal{B}_s$ contains every distinct feasible value of $b_i(k)$ over all devices and layer counts in this range (lines 3-4).
We set $b_i(k)=\infty$ when the corresponding memory constraint is infeasible.
The largest device cost of any complete allocation must equal one value in $\mathcal{B}_s$.
For each $\tau\in\mathcal{B}_s$, the scheduler therefore considers allocations satisfying $b_i(k)\leq\tau$ on every device (lines 6-13).
Let $F_{\tau}(l,i)$ be the minimum computation time for assigning the first $l$ layers to the first $i$ devices under this constraint.
Only states satisfying $is\leq l\leq|\mathcal{L}|-(|\mathcal{D}|-i)s$ need to be instantiated because every processed and unprocessed device must receive at least $s$ layers.
We set $F_{\tau}(0,0)=0$ and all infeasible states to infinity.
The state transition is
\begin{equation}
\label{eq:state_transition}
F_{\tau}(l,i)=
\min_{\substack{s\leq k\leq l-(i-1)s\\ b_i(k)\leq\tau}}
\left\{F_{\tau}(l-k,i-1)+k\cdot\textit{comp}(\{\ell\})\right\}.
\end{equation}
For each reachable state, $P_{\tau}(l,i)$ stores the selected layer count $k$ of device $i$.
To recover an allocation, backtracking starts from $(l,i)=(|\mathcal{L}|,|\mathcal{D}|)$, assigns $k_i=P_{\tau}(l,i)$, and moves to $(l-k_i,i-1)$ until it reaches $(0,0)$.
When candidate costs tie, the scheduler stores the larger $k$ to make the result deterministic without changing the objective value.
Thresholds are processed sequentially, so the same predecessor table is reused and only the allocation of the best threshold is retained.
The scheduler selects the threshold (lines 14-16) and allocation that minimize
\begin{equation}
\label{eq:threshold_latency}
T_s=\min_{\tau\in\mathcal{B}_s}
\max\left\{F_{\tau}(|\mathcal{L}|,|\mathcal{D}|)+T_{comm},\tau\right\}.
\end{equation}
The retained backtracking result gives the number of layers assigned to each device.
We then balance both the total number of layers and the number of offloaded layers across stages, assigning larger units to larger stages, and determine the layer identities according to the pipeline execution order (lines 18).
A stage count is infeasible when every threshold produces an infinite terminal state.

\begin{theorem}\label{thm:allocation_optimality}
Under the stated layer-cost and timing assumptions with zero initial KV cache occupancy, the DP phase minimizes modeled latency over whole-layer allocations for each feasible $s$.
\end{theorem}
\begin{proof}
Fix a feasible stage count $s$.
Suppose that device $i$ is assigned $k$ layers.
Equation~\ref{eq:unit_memory} defines $q_i(k)$ as the minimum number of layers that must be offloaded to satisfy the memory constraint.
Under the assumption that all layers incur identical computation and loading costs, the computation and loading terms contributed by the offloaded weights to $b_i(k)$ are nondecreasing in the number of offloaded layers.
Therefore, $q_i(k)$ minimizes $b_i(k)$ among all memory-feasible whole-layer placements with the same $k$.
Balancing these offloaded layers across the $s$ stages realizes the largest loading unit $\lceil q_i(k)/s\rceil$ used in Eq.~\ref{eq:unit_memory}.

For any $\tau\in\mathcal{B}_s$, we show by induction on $i$ that $F_{\tau}(l,i)$ is the minimum computation time among all feasible assignments of the first $l$ layers to the first $i$ devices with device costs bounded by $\tau$.
The claim holds for the initial state $F_{\tau}(0,0)=0$.
For the inductive step, every feasible assignment places some $k$ layers on device $i$ and leaves a feasible assignment of $l-k$ layers to the preceding devices.
Equation~\ref{eq:state_transition} examines every such $k$ and selects the minimum computation time.
Thus, the claim holds for all reachable states.

Let $\mathcal{A}^{*}$ be an optimal whole-layer allocation and let $\tau^{*}=\max_i b_i(k_i^{*})$ be its largest device cost.
Because $\tau^{*}\in\mathcal{B}_s$, the DP evaluates the constraint defined by $\tau^{*}$.
The allocation $\mathcal{A}^{*}$ is feasible for this constraint, so $F_{\tau^{*}}(|\mathcal{L}|,|\mathcal{D}|)$ cannot exceed its total computation time.
Equation~\ref{eq:threshold_latency} therefore returns an objective value no greater than the modeled latency of $\mathcal{A}^{*}$.
The backtracked solution is itself feasible and its largest device cost is no greater than its selected threshold.
Its modeled latency cannot exceed the value returned by Eq.~\ref{eq:threshold_latency}.
Since no feasible allocation can improve upon $\mathcal{A}^{*}$, the returned allocation has the same minimum modeled latency.
\end{proof}

\textbf{Lightweight block refinement.}
After the DP-based allocation algorithm, a bottleneck device may still exist, where the loading time of this device becomes sufficiently long to constrain the overall system loading process.
To address this issue, we further explore a Lightweight block refinement (lines 20-29).
The remaining memory of the bottleneck device may fit a block even when it cannot fit another whole layer.
Starting from the DP solution of each stage configuration, the scheduler identifies the device with the largest $b_i$ and attempts to retain the MLP (multi-layer perceptron) block of one fully offloaded layer.
If this conversion is infeasible, it further attempts to retain the MHA (multi-head attention) block.
These conversions reduce the loading cost by $p_M\cdot\textit{load}(\{\ell\})$ and $p_A\cdot\textit{load}(\{\ell\})$, respectively.
Each candidate conversion is accepted only if the resident weight footprint plus the largest actual loading unit remains within $Mem_i$.
The reserved buffer is released only when the corresponding largest loading unit is reduced.

After each successful conversion, $b_i$ is recomputed based on the loading and computation costs of the remaining offloaded blocks.
The refinement terminates when the bottleneck device cannot accept any further conversion.
Since the total computation and communication costs remain unchanged and the refinement can only reduce or preserve $b_i$, Eq.~\ref{eq:equivalent_latency} ensures that the modeled latency does not increase.
Finally, the scheduler compares all stage candidates after refinement and returns the best configuration.
The obtained solution is guaranteed to be no worse than the whole-layer optimum.

\textbf{Complexity Analysis.}
Let $L=|\mathcal{L}|$, $D=|\mathcal{D}|$, $S=\lfloor L/D\rfloor-1$, and $R_s=|\mathcal{B}_s|$.
For fixed $s$ and $\tau$, the DP instantiates $O(DL)$ states and examines at most $L$ layer counts per state, giving $O(DL^2)$ time.
Processing all thresholds for $s$ therefore takes $O(R_sDL^2)$ time.
The direct precomputation of $q_i(k)$ takes at most $O(DL^2)$ time, and threshold construction and sorting take $O(DL\log(DL))$ time.
With a direct scan of the offloaded layers, block refinement takes $O(L^2+L\log D)$ time because there are at most $L$ successful conversions, each scan examines at most $L$ layers, and each iteration updates the device heap.
Across all $S$ candidate stage counts, the total running time is
\begin{equation}
\label{eq:scheduler_complexity}
O\!\left(\sum_{s=2}^{\lfloor L/D\rfloor}
\left(R_sDL^2+DL^2+DL\log(DL)\right)\right).
\end{equation}
Because $s\leq k\leq L-(D-1)s$, we have $R_s\leq D(L-Ds+1)\leq DL$.
Since $S<L/D$, the total scheduler time $T_{\mathrm{sched}}$ has the following conservative worst-case bound:
\begin{equation}
\label{eq:scheduler_upper_bound}
O\!\left(\frac{L}{D}\cdot D^2L^3\right)
=O(DL^4).
\end{equation}

In practice, the scheduler performs fewer updates because infeasible layer counts are excluded, duplicate values of $b_i(k)$ produce only one threshold, and the constraint $b_i(k)\leq\tau$ filters transitions for each threshold.
For LLaMA3.3-70B-Instruct, which has 80 layers, the scheduler completes within 2 seconds for the four-device configuration.

\subsection{Online Memory Adaptation Strategy}\label{online}
During inference, KV cache growth consumes additional device memory, while network bandwidth fluctuates over time.
Repeatedly running the allocation scheduler and reallocating the resident and offloaded layers for each device to handle these changes would incur excessive overhead.
To handle these changes, we propose an online memory adaptation strategy with an online memory-aware planner and a bandwidth-sensitive KV cache transfer protocol.

\textbf{Online memory-aware planner.}
As the KV cache accumulates, the planner triggers progressive offloading at predefined thresholds $TS^j_i$ for the device $i$.
The first threshold is derived from the remaining available memory after layer allocation:
\begin{equation}\label{eq:first_threshold}
TS_i^1=\max\left\{n\in\mathbb{Z}:\,
M_i^{\mathrm{KV}}(n)\leq Mem_i-M_i^{W}\right\}.
\end{equation}
This threshold includes the maximum amount of KV cache required by the i-th device during a specific stage, as well as any KV cache that might be transferred from other devices.
The planner activates the next offloading plan before further token growth exceeds this memory budget.

To minimize additional loading latency, let $\alpha$ and $\beta$ count additional MHA and MLP blocks offloaded per stage relative to the initial allocation.
Sharing their loading space across stages releases $(\#Stage-1)$ times the added unit size.
For the next required token count $n$, the planner minimizes additional per-step loading:
\begin{equation}
\label{eq:threshold_constrain}
\begin{aligned}
\min_{\alpha,\beta}\quad
&\#Stage\cdot \textit{load}(\{\ell\}) \cdot (\alpha p_A+\beta p_M)\\
\text{s.t.}\quad
&M_i^{\mathrm{KV}}(n)\leq Mem_i-M_i^{W}\\
&\qquad +(\#Stage-1) \cdot (\alpha p_A+\beta p_M) \cdot l_{\mathrm{size}},\\
&\alpha,\beta\in\mathbb{Z}_{\geq0}.
\end{aligned}
\end{equation}

The planner enumerates only combinations supported by the resident blocks available in every stage under the initial allocation.
The same offloading plan is applied to all stages within one device.
Since loading across devices is mutually overlapped in the interleaved pipeline, only a single loading overhead is incurred.
When threshold $TS^j_i$ is reached, the corresponding plan is triggered.
For instance, at $TS^1_i$ the MHA block is offloaded, and at $TS^2_i$ the MLP block is offloaded while the previously evicted MHA block is reloaded.
The switching overhead is bounded by a single block (MHA or MLP), which is negligible relative to the total loading volume of the interleaved pipeline parallelism.


\textbf{Network bandwidth-sensitive KV cache transfer protocol.}
Due to device heterogeneity, some devices may trigger offloading prematurely and become bottlenecks. 
To mitigate this, each low-threshold device transfers part of its KV cache to a paired high-threshold device $d_{target}$, thereby delaying its next offloading threshold. 

The signed transfer-time margin for the KV cache is
\begin{equation}
\label{eq:signed_time_budget}
\Delta_i =
\textit{load}(\Tilde{\mathcal{L}}_i)
- T_{\mathrm{comm}}
- \sum_{i' \in \mathcal{D}-\{i\}}
  \textit{comp}(\mathcal{L}_{i'})
- \textit{comp}
  (\mathcal{L}_i-\Tilde{\mathcal{L}}_i).
\end{equation}

Then, the KV cache calculation formula for device $i$ on the stage $s$ is used to calculate how much memory space each token location needs to occupy. The formula is as follows:
\begin{equation}
\label{eq:kv_bytes_per_token}
\kappa_{i,s}
=
2 z b
\sum_{\ell \in \mathcal{L}_{i,s}}
H_{\ell}^{\mathrm{KV}} d_{\ell}.
\end{equation}
The factor 2 accounts for the key and value caches. Here, $z$ is the number of bytes per KV element ($z=2$ for bfloat16), $b$ is the
inference micro-batch size, $H_{\ell}^{\mathrm{KV}}$ is the number of KV heads, and $d_{\ell}$ is the head dimension.

Let $\kappa_i \triangleq \sum_{s=1}^{\#Stage}\kappa_{i,s}$ denote the per-token KV cache footprint across all stages on device $i$.
The signed token budget for KV cache transfer is estimated as
\begin{equation}
\label{eq:signed-transfer-tokens}
V_i^{\mathrm{trans}}
=
\operatorname{sgn}(\Delta_i)
\left\lfloor
\frac{bw_{\mathrm{net}}|\Delta_i|}{\kappa_i}
\right\rfloor,
\qquad
V_i^{\mathrm{trans}}\in\mathbb{Z}.
\end{equation}

Alg.~\ref{algo:trans_strategy} details the transfer procedure.
For each stage on the device $i$, \workname{} transfers part of its KV cache after execution and fetches it back before the stage is executed again, reducing its memory pressure.
A positive $V_i^{trans}$ denotes an outgoing budget, while a negative value denotes a receiving budget of magnitude $|V_i^{trans}|$.
The actual transfer is limited by both devices' budgets, available KV entries, and the receiver's available memory.

To adapt to bandwidth fluctuations without costly re-allocation, \workname{} monitors the current bandwidth $bw'_{net}$ before each step and recomputes $(V_i^{trans})'$ via Eq.~\ref{eq:signed-transfer-tokens}. 
When bandwidth decreases, the transfer volume is immediately reduced to avoid extra waiting delays. 
When bandwidth increases, the adjustment is deferred unless the next offloading threshold $TS^{j+1}_i$ is imminent, thereby avoiding unnecessary modification overhead from frequent bandwidth change (lines 10-12).
A fluctuation threshold $n_{ts}$ is employed to filter out minor variations.

\begin{algorithm}[t]
  \small
  \caption{Online KV Cache Transfer Protocol}
  \label{algo:trans_strategy}
  \SetKwInput{Input}{Input}
  \SetKwInput{Output}{Output}

  \Input{Number of KV cache tokens to transfer per stage for the device $i$ $V_i^{trans}$, target device for KV cache transfer $d_{target}$, network bandwidth $bw_{net}$, KV cache of the device $i$ at each stage $cache_s$}
  \Output{Generated sequence}
  \For{$s=1$ \KwTo $\#Stage$}{
    Complete forward propagation for this stage\\
    \If{$V_i^{trans} > 0$}{
          Transfer $cache_s[-V_i^{trans}:]$ to $d_{target}$\\
          Delete $cache_s[-V_i^{trans}:]$\\
        }
  }
  Initialize $cache\_ready_s$ for each stage\\
  \For{sequence length $n = 1$ \KwTo  $max\_length$}{
        Monitor the network bandwidth $bw_{net}'$\\
        Get $(V_i^{trans})'$ by Eq.~\ref{eq:signed-transfer-tokens}\\
        \If{$|(V_i^{trans})' - V_i^{trans}|\geq n_{ts}$}{
    \If{$(V_i^{trans})' \leq V_i^{trans}$ or
        $n + V_i^{trans} \geq TS^{j+1}_i - 1$}{
        $V_i^{trans}\gets(V_i^{trans})'$\\
    }
}
        \For{$s=1$ \KwTo $\#Stage$}{
            Wait for $cache\_ready_s$\\
            $cache_s \gets$ concat $(cache_s,cache_{trans})$\\
            Complete forward propagation for this stage\\
            \If{$(V_i^{trans}) > 0$}{
              Transfer $cache_s[-(V_i^{trans}):]$ to $d_{target}$\\
              Delete $cache_s[-(V_i^{trans}):]$\\
              Asynchronously receive $cache_{(s \bmod \#Stage)+1}$\\
            }
        }
        \If{generated sequence finished}{
            \Return generated sequence\\
        }
    }
  \Return generated sequence\\
\end{algorithm}

\section{Experiments}
\label{exp}
\subsection{Experimental Settings}\label{exp:settings}
\textbf{Testbed Implementation.}
Our testbed consists of three types of NVIDIA Jetson developer kits, whose specifications are listed in Tab.~\ref{equipment}.
The devices are interconnected via a router and a switch at 1000~Mbps, and we use the Linux TC tool~\cite{hubert2002linux} to control bandwidth for constrained-network experiments.
\workname{} is implemented with \textbf{2500+} lines of Python and \textbf{500+} lines of C++/CUDA.
We evaluate under sporadic and bursty request patterns common in edge deployments.
The \emph{sporadic} pattern processes one active sequence with micro-batch size 1, whereas the \emph{bursty} pattern processes $|\mathcal{D}|$ concurrent sequences with micro-batch size $|\mathcal{D}|$.

\textbf{Baselines.}
We compare \workname{} against 6 baselines including pipeline parallelism (PP) and tensor parallelism (TP).
For baselines that do not natively support memory-constrained execution,
we evict KV cache entries that exceed the available memory and recompute
those entries during subsequent forward passes.

\begin{itemize}
    \item \textbf{Pipeline parallelism}~\cite{huang2019gpipe} partitions model layers across devices according to memory capacity and computing capability.
    \item \textbf{Pipeline + offloading} extends pipeline parallelism with dynamic layer offloading, where a portion of model layers is offloaded to storage to ensure that the full model and the required KV cache can be accommodated under tight memory budgets.
	\item \textbf{EdgeShard}~\cite{zhang2024edgeshard} explicitly models heterogeneous device capabilities and employs dynamic programming to optimize model partitioning for edge PP inference.
	\item \textbf{Galaxy}~\cite{ye2024galaxy} combines fine-grained workload partitioning with sequence parallelism for TP-based distributed inference.
	\item \textbf{TPI-LLM}~\cite{li2025tpi} employs sliding-window memory management and communication link optimization for TP inference on memory-constrained edge devices.
    \item \textbf{TPI-LLM+offloading} extends TPI-LLM with a larger sliding window instead of recomputation to handle KV cache memory pressure.
\end{itemize}

Within each panel, all methods use the same model checkpoint, fixed input/output lengths, device subset, device power modes, and TC-controlled link rate.
The reported variants differ in their partitioning, parallel execution, and memory-management mechanisms described above.
Considering that Jupiter~\cite{ye2025jupiter} introduces an additional speculative decoding head, it incurs extra memory consumption.
Moreover, Jupiter assumes that the entire model remains resident in memory during inference.
Therefore, we do not include Jupiter as a baseline in our evaluation, since its memory requirements and system assumptions differ from those of our target scenario.

\begin{table}[]
\vspace{-1em}
\caption{The specifications of the experimental equipment in detail.}
\label{equipment}
\centering
\renewcommand{\arraystretch}{1.2}
\scalebox{0.8}{
\begin{tabular}{ccccc}
\toprule
{Type}&{Memory}&{GPU}&{AI Performance}&{Power Mode} \\
\midrule
\makecell[c]{Jetson Xavier\\NX 16GB} &16GB &\makecell[c]{384-core NVIDIA Volta™\\ architecture GPU with \\ 48 Tensor Cores}&21TOPS &20W\\ 
\makecell[c]{Jetson AGX\\ Orin 32GB} &32GB &
\makecell[c]{1792-core NVIDIA Ampere\\ architecture GPU with \\ 56 Tensor Cores }&200TOPS &50W\\
\makecell[c]{Jetson AGX\\ Orin 64GB} &64GB &\makecell[c]{2048-core NVIDIA Ampere\\ architecture GPU with \\ 64 Tensor Cores}&275TOPS &60W\\
\bottomrule
 \end{tabular}}
\end{table}

\begin{figure}[]
    \centering
    \includegraphics[width=0.8\linewidth]{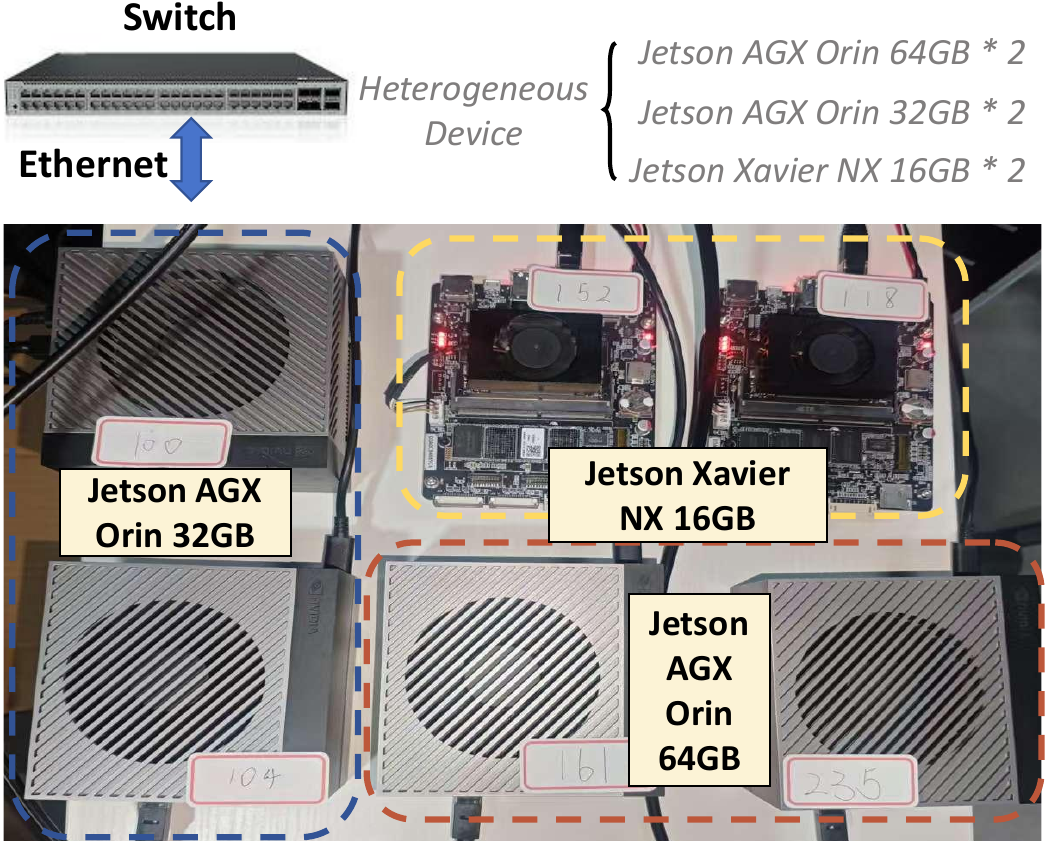}
    \caption{Overview of the main testbed platform, including 3 types of Jetson devices.}
    \label{fig:exp_device}
\end{figure}

\textbf{Models.}
We evaluate on three models spanning 13B to 70B parameters: LLaMA2-13B-Instruct, Qwen3-32B~\cite{yang2025qwen3}, and LLaMA3.3-70B-Instruct~\cite{dubey2024LLaMA}, all downloaded from Hugging Face.
We use a subset of a general-purpose text generation dataset WikiText-2~\cite{merity2016pointer} with fixed-length inputs and outputs, following EdgeShard~\cite{zhang2024edgeshard}.
We configure all methods to generate the same fixed number of output tokens with the same prompt.
Once the KV cache exhausts the available GPU memory, the system enters the memory-saturated regime, where subsequent tokens require explicit offloading or KV recomputation to complete the forward pass.

\textbf{Evaluation Metrics.}
We report per-token inference latency (ms/token) under both sporadic and bursty request patterns.
Since \workname{} does not involve any modification to the model parameters, the reported results only evaluate the decoding-stage performance of different methods on the original model.
Under the bursty request pattern, decoding efficiency is calculated as the generation latency per token.
Some baselines are marked as OOM (out-of-memory) because they fail to allocate model slices under insufficient device memory.
Some baselines are marked as OOT (out-of-time) when they can complete inference but fail to meet the practical latency requirement for LLM inference.
Specifically, we classify a method as OOT if its latency exceeds 5s/token under the sporadic request pattern or 2s/token under the bursty request pattern.

\textbf{Experimental Setups.}
We configure two groups of experiments.
For scenarios with \textit{severe memory constraints}, we use 1 $\times$ \textit{Jetson AGX Orin 64GB}, 3 $\times$ \textit{Jetson AGX Orin 32GB}, and 1 $\times$ \textit{Jetson Xavier NX 16GB}, with three increasingly stringent memory constraints: (i) no memory reservation, (ii) 3\,GB reserved on each of two \textit{Orin 32GB} devices, and (iii) an additional 5\,GB on the \textit{Orin 64GB} and 1\,GB on the \textit{NX 16GB}.
For scenarios with \textit{KV cache memory constraints}, we use three device configurations matched to model scales: the 13B model on 1 $\times$ \textit{Orin 32GB} + 1 $\times$ \textit{NX 16GB}; the 32B model on 1 $\times$ \textit{Orin 64GB} + 1 $\times$ \textit{Orin 32GB} + 1 $\times$ \textit{NX 16GB}; and the 70B model on 2 $\times$ \textit{Orin 64GB} + 1 $\times$ \textit{Orin 32GB} + 1 $\times$ \textit{NX 16GB}.
All configurations are tested under both 100\,Mbps and 200\,Mbps bandwidth settings.

\begin{figure*}[t]
    \centering
    \includegraphics[width=\linewidth]{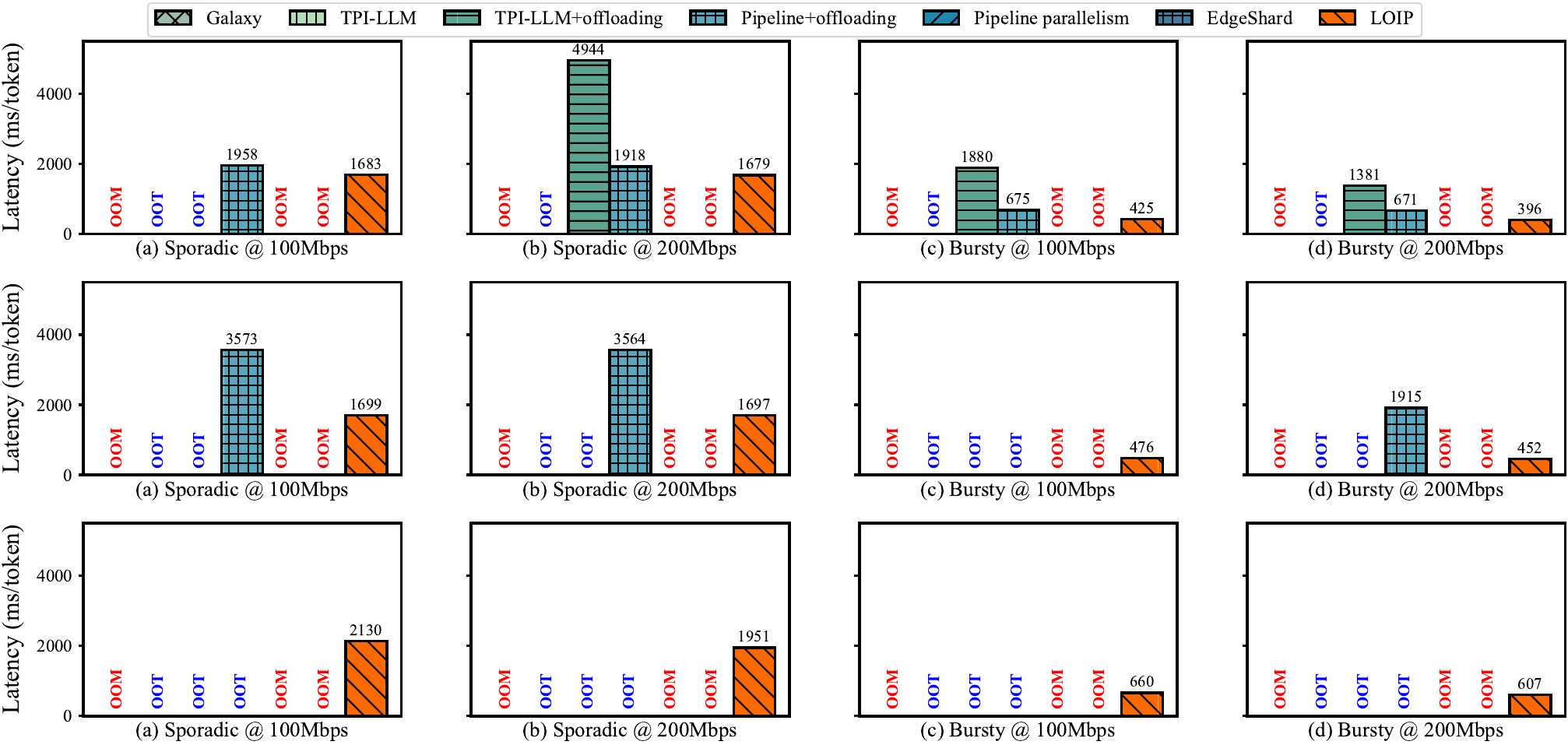}
    \caption{Decoding efficiency, measured as per-token decoding latency, of \workname{} and baselines on LLaMA3.3-70B-Instruct under three progressively severe memory constraints (top to bottom): no reservation, partial reservation, and further reservation. For each budget, we report results under 100\,Mbps and 200\,Mbps bandwidth with both sporadic and bursty request patterns.}
    \label{fig:70_scarce}
\end{figure*}

\subsection{Decoding Efficiency under Severe Memory Constraints}
We evaluate \workname{} on LLaMA3.3-70B-Instruct under three progressively increased memory reservation configurations on the same edge devices.
Its decoding latency is compared with all baselines under two bandwidth settings and diverse request patterns, as shown in Fig.~\ref{fig:70_scarce}.

\textbf{Comparison with TP-based parallel inference methods.}
TP-based methods are highly sensitive to extremely memory-constrained edge settings.
Galaxy consistently fails with OOM because it assumes that each device can hold its assigned TP slice, which no longer holds when effective device memory is further reduced. 
Although TPI-LLM and TPI-LLM+offloading reduce model memory pressure through offloading, their TP execution still requires frequent cross-device synchronization and communication.
Under sporadic request patterns, such overhead cannot be sufficiently amortized, causing them to exceed the practical latency threshold in most settings.
Under bursty request patterns, larger micro-batches improve feasibility only in less constrained cases, while latency quickly increases or becomes OOT as memory becomes tighter. 
In contrast, \workname{} completes inference across all evaluated low-memory and bandwidth-constrained settings.
This shows that directly combining TP with KV cache recomputation or weight offloading remains insufficient for full-parameter decoding on the evaluated bandwidth-limited edge devices.
\workname{} avoids this limitation through offloading-based interleaved pipeline parallelism, which reduces synchronization overhead, and overlaps model loading with computation and communication.

\textbf{Comparison with PP-based parallel inference methods.}
Pipeline parallelism and EdgeShard are marked as OOM in these settings because they require devices to keep their assigned model slices in memory. 
Pipeline+offloading improves memory feasibility by dynamically loading model slices, but a substantial portion of the loading latency is not effectively overlapped, especially when more layers must be offloaded under tighter memory budgets. 
As a result, it either incurs much higher latency than \workname{} or becomes OOT as the memory constraint becomes more severe.
In contrast, \workname{} maintains relatively low inference latency.
Its UMA-aware loading optimization lowers the latency of dynamically loading offloaded layers on edge devices.
Meanwhile, the fine-grained allocation scheduler improves the assignment of execution layers and offloaded layers, enabling \workname{} to better exploit heterogeneous device resources.

\begin{figure*}[t]
    \centering
    \includegraphics[width=\linewidth]{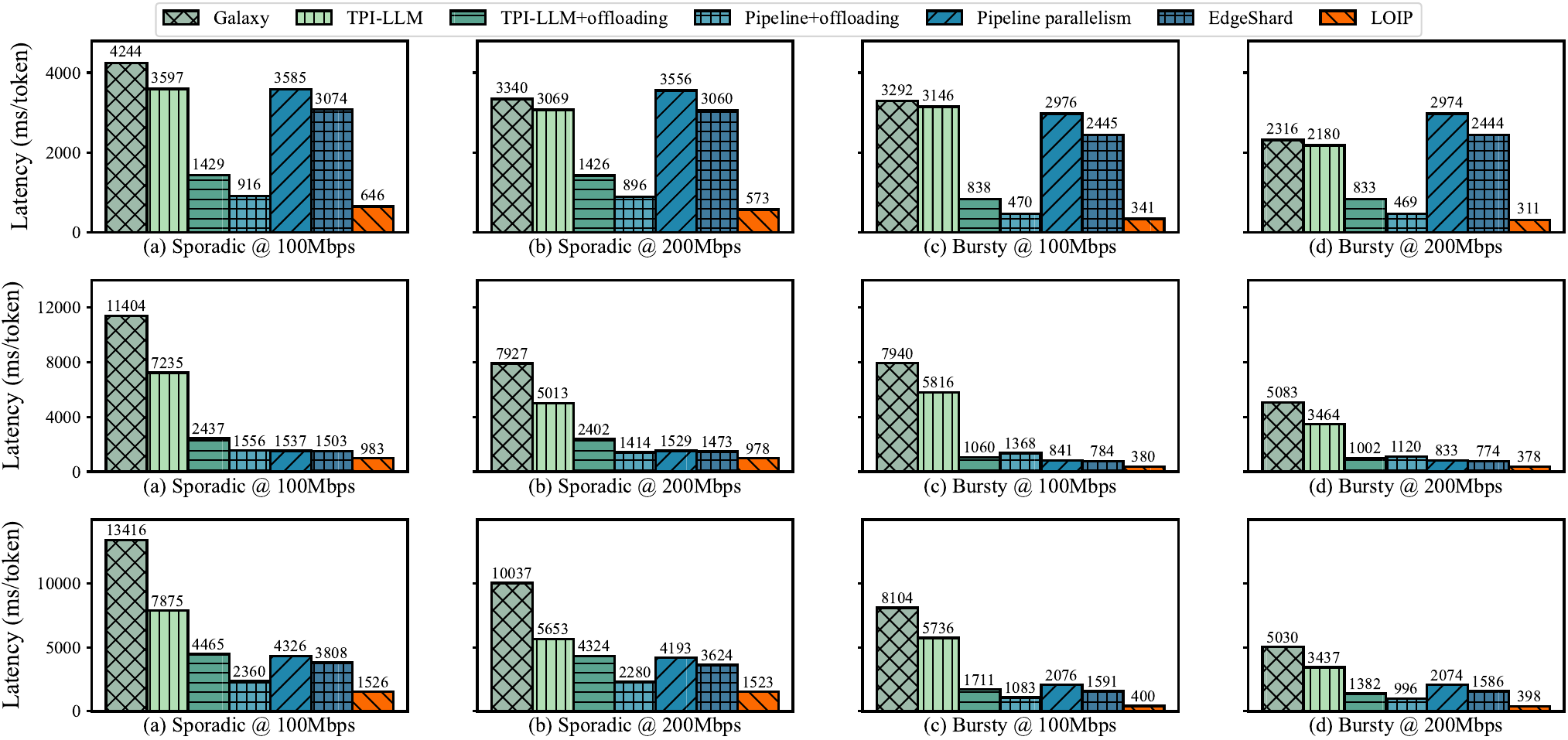}
    \caption{Decoding efficiency of \workname{} and the baselines under KV cache memory constraints across three model scales (top to bottom): LLaMA2-13B-Instruct, Qwen3-32B, and LLaMA3.3-70B-Instruct. For each model, we report results under 100\,Mbps and 200\,Mbps bandwidth with both sporadic and bursty request patterns.}
    \label{fig:main}
\end{figure*}

\subsection{Decoding Efficiency under KV Cache Memory Constraints}
In this section, we evaluate the inference decoding efficiency of \workname{} under two bandwidth settings and two request patterns, as shown in Fig.~\ref{fig:main}.
Unlike the extremely memory-constrained scenarios, the heterogeneous devices in this section are configured with sufficient memory to accommodate their assigned model slices, so that no baseline fails during the initial model placement. 
This setting allows us to focus on the overall inference latency of different collaborative execution strategies under the runtime memory pressure caused by KV cache growth.

\textbf{Comparison with TP-based parallel inference methods.}
TP-based methods generally suffer from high latency in bandwidth-limited edge environments. 
Although Galaxy, TPI-LLM, and TPI-LLM+offloading can execute inference when memory is sufficient for them to hold the model slices, their TP execution still requires frequent synchronization and communication across devices. 
This overhead becomes more pronounced under low-bandwidth settings, leading to substantially higher latency than \workname{}. 
Compared with these TP-based baselines, \workname{} achieves 2.5$\times$$\sim$20.9$\times$ and 2.7$\times$$\sim$13.4$\times$ speedups under bursty requests at 100 Mbps and 200 Mbps, respectively. 
Under sporadic requests, \workname{} achieves 2.2$\times$$\sim$11.6$\times$ and 2.5$\times$$\sim$8.1$\times$ speedups under 100 Mbps and 200 Mbps, respectively. 
These results indicate that even when memory is sufficient to place the model, TP-based inference remains inefficient on bandwidth-limited edge devices, whereas \workname{} reduces synchronization overhead and further hides loading latency with its offloading-based interleaved pipeline parallelism.

\begin{figure}[t]
    \centering
    \includegraphics[width=\linewidth]{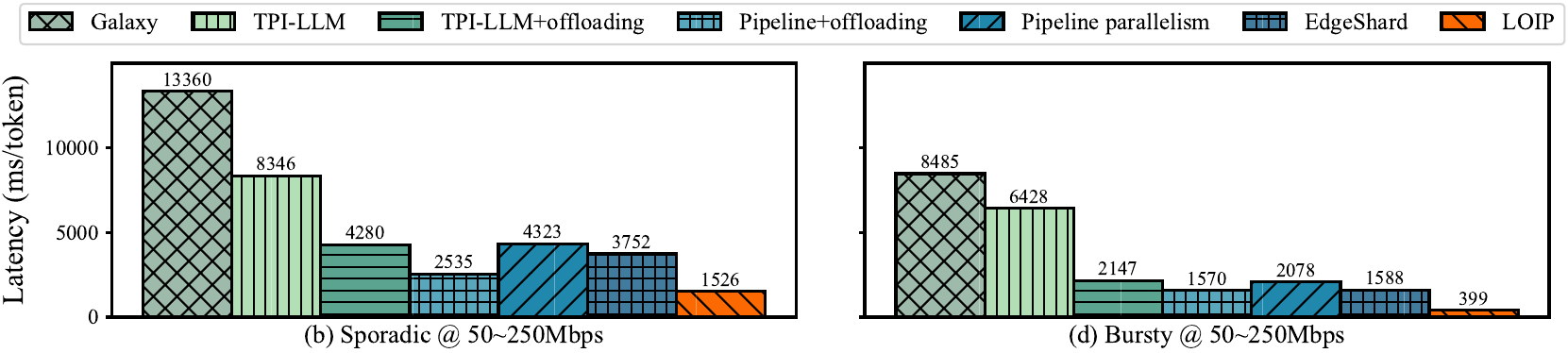}
    \caption{Decoding efficiency comparison under varying network bandwidth on the LLaMA3.3-70B-Instruct model.}
    \label{fig:vary_draw}
\end{figure}

\begin{table*}[h]
\caption{Effectiveness of \workname{} Components.}
\label{ablation}
\centering
\renewcommand{\arraystretch}{0.95}
\resizebox{\textwidth}{!}{%
\begin{tabular}{lcccc}
\toprule
\multirow{2}{*}{Methods}
& \multicolumn{2}{c}{Sporadic}
& \multicolumn{2}{c}{Bursty} \\
\cmidrule(lr){2-3}\cmidrule(lr){4-5}
& Latency (ms/token) & \workname{} Speedup
& Latency (ms/token) & \workname{} Speedup \\
\midrule
\workname{} without KV cache transfer protocol & 1614.2 & 1.05$\times$ & 455.6 & 1.12$\times$ \\
\workname{} without fine-grained allocation scheduler   & 1939.8 & 1.26$\times$ & 482.2 & 1.18$\times$ \\
\workname{} without UMA-aware loading optimization    & 2251.7 & 1.47$\times$ & 488.4 & 1.20$\times$ \\
\workname{}                                    & 1533.9 & 1.00$\times$ & 407.1 & 1.00$\times$ \\
\bottomrule
\end{tabular}}
\end{table*}

\textbf{Comparison with PP-based parallel inference methods.}
Pipeline parallelism and EdgeShard still rely on static layer placement and cannot effectively handle the runtime memory pressure introduced by KV cache growth. 
Although pipeline+offloading maintains relatively good performance under the sporadic request pattern, it suffers from significantly higher latency under the bursty request pattern.
This is because concurrent requests enlarge loading bubbles, causing devices to spend more time waiting for required layers.
In contrast, \workname{} achieves 1.4$\times$$\sim$8.7$\times$ and 1.5$\times$$\sim$9.6$\times$ speedups under bursty requests, and achieves 1.4$\times$$\sim$5.5$\times$ and 1.4$\times$$\sim$6.2$\times$ speedups under sporadic requests at 100 Mbps and 200 Mbps, respectively.
The improvement mainly comes from the offloading-based interleaved pipeline parallelism and the UMA-aware loading optimization.
The former reduces the loading bubbles during inference, while the latter further reduces the latency of dynamically loading offloaded layers on edge devices.
In addition, the online memory adaptation strategy alleviates latency degradation caused by KV cache growth, thereby preventing individual devices from becoming performance bottlenecks in the overall inference system.

\subsection{Decoding efficiency under Varying Network Bandwidth}
In this section, we evaluate the robustness of \workname{} under dynamic network bandwidth variations.
We conduct the experiment on the LLaMA3.3-70B-Instruct model using a heterogeneous device configuration consisting of 2 \textit{Jetson AGX Orin 64GB} devices, 1 \textit{Jetson AGX Orin 32GB} device, and 1 \textit{Jetson Xavier NX 16GB} device.
As shown in Fig.~\ref{fig:vary_draw}, we compare the inference latency of \workname{} against all baselines under both sporadic and bursty request patterns.
To emulate time-varying edge network conditions, we generate a sequence of bandwidth-switching intervals measured by the number of generated tokens.
When the generated token count reaches the next switching interval, we randomly sample a new bandwidth value between 50~Mbps and 250~Mbps and apply it to the inter-device links.

\workname{} consistently outperforms all baselines under dynamic bandwidth variations.
Under the bursty request pattern, \workname{} achieves 3.9$\times$$\sim$21.3$\times$ speedups over the baseline methods.
Under the sporadic request pattern, \workname{} achieves 1.7$\times$$\sim$8.8$\times$ speedups.
These results demonstrate that \workname{} can maintain efficient collaborative inference even when network bandwidth changes at runtime.
The online bandwidth-aware KV cache transfer protocol exploits idle bandwidth to transfer generated KV cache based on bandwidth variations, keeping memory usage balanced across devices.
This prevents excessive memory consumption on individual devices from triggering additional layer offloading and increasing the overall system latency.

\subsection{Ablation Study}
To explore the effectiveness of each component in our \workname{} framework, we design three ablation experiments on the LLaMA3.3-70B-Instruct model with 2 \textit{Jetson AGX Orin 64GB} devices, 1 \textit{Jetson AGX Orin 32GB} device, and 1 \textit{Jetson Xavier NX 16GB} device. 
The ablation experiment setup is as follows:

$\bullet$ \textbf{\workname{} without KV cache transfer protocol:}
To investigate the effectiveness of the KV cache transfer protocol, we disable it in this experimental setup, preventing any transfer of KV cache.

$\bullet$ \textbf{\workname{} without fine-grained allocation scheduler:}
This variant disables the fine-grained allocation scheduler and replaces it with a uniform layer allocation strategy to assign consecutive model layers to devices under memory constraints.

$\bullet$ \textbf{\workname{} without UMA-aware loading optimization:}
This variant removes the UMA-aware loading optimization while keeping the same offloading-based interleaved pipeline parallelism.

Table~\ref{ablation} shows the marginal effect of removing one component in this device configuration. Disabling KV cache transfer raises latency from 1533.9 to 1614.2\,ms/token in the sporadic regime (5.2\%) and from 407.1 to 455.6\,ms/token in the bursty regime (11.9\%).
Replacing the scheduler with uniform allocation raises latency by 26.5\% and 18.4\%, respectively, while removing UMA-aware loading raises it by 46.8\% and 20.0\%.
The KV cache transfer protocol uses idle network bandwidth to move KV cache from devices with limited free memory to peers with more available capacity.
This redistribution delays additional layer offloading on memory-constrained devices and helps prevent any single device from becoming the system's loading bottleneck.
The fine-grained allocation scheduler uses the cost model to assign execution and offloaded layers according to device heterogeneity, memory budgets, loading overhead, and communication costs.
Removing it leads to imbalanced pipeline execution and higher decoding latency.
Finally, the UMA-aware loading optimization reduces the runtime overhead of frequent model slice loading during execution.
Without this optimization, offloaded model slices fall back to the conventional loading path, adding CPU-side materialization, duplicate data movement, and synchronization overhead.

\section{Conclusion}
In this paper, we presented \workname{}, an offloading-based collaborative LLM inference framework for heterogeneous, memory-constrained UMA-based edge devices without model compression.
\workname{} combines model offloading with interleaved pipeline parallelism to overlap weight loading with computation and inter-device communication.
Its cost model and fine-grained scheduler select layer assignments, offloading decisions, and stage counts, while UMA-aware loading reduces repeated model-slice loading overhead.
The online adaptation strategy further adjusts block-level offloading and KV cache placement as memory pressure and bandwidth change.
Experiments with 13B--70B models on heterogeneous NVIDIA Jetson testbeds show up to $8.8\times$ and $20.3\times$ speedups over the evaluated baselines under single-input and concurrent-input execution, respectively, while leaving the original model parameters unchanged.

\bibliographystyle{IEEEtran}
\bibliography{example_paper-ToN}

@article{zhang2024edgeshard,
  title={Edgeshard: Efficient llm inference via collaborative edge computing},
  author={Zhang, Mingjin and Shen, Xiaoming and Cao, Jiannong and Cui, Zeyang and Jiang, Shan},
  journal={IEEE Internet of Things Journal},
  year={2024},
  publisher={IEEE}
}

@inproceedings{ye2025jupiter,
  title={Jupiter: Fast and resource-efficient collaborative inference of generative llms on edge devices},
  author={Ye, Shengyuan and Ouyang, Bei and Zeng, Liekang and Qian, Tianyi and Chu, Xiaowen and Tang, Jian and Chen, Xu},
  booktitle={IEEE INFOCOM 2025-IEEE Conference on Computer Communications},
  pages={1--10},
  year={2025},
  organization={IEEE}
}

@inproceedings{ye2024galaxy,
  title={Galaxy: A resource-efficient collaborative edge ai system for in-situ transformer inference},
  author={Ye, Shengyuan and Du, Jiangsu and Zeng, Liekang and Ou, Wenzhong and Chu, Xiaowen and Lu, Yutong and Chen, Xu},
  booktitle={IEEE INFOCOM 2024-IEEE Conference on Computer Communications},
  pages={1001--1010},
  year={2024},
  organization={IEEE}
}

@article{li2025tpi,
  title={TPI-LLM: Serving 70B-scale LLMs Efficiently on Low-resource Mobile Devices},
  author={Li, Zonghang and Feng, Wenjiao and Guizani, Mohsen and Yu, Hongfang},
  journal={IEEE Transactions on Services Computing},
  year={2025},
  publisher={IEEE}
}

@inproceedings{kwon2023efficient,
  title={Efficient memory management for large language model serving with {PagedAttention}},
  author={Kwon, Woosuk and Li, Zhuohan and Zhuang, Siyuan and Sheng, Ying and Zheng, Lianmin and Yu, Cody Hao and Gonzalez, Joseph and Zhang, Hao and Stoica, Ion},
  booktitle={Proceedings of the 29th Symposium on Operating Systems Principles},
  pages={611--626},
  year={2023}
}

@article{huang2019gpipe,
  title={Gpipe: Efficient training of giant neural networks using pipeline parallelism},
  author={Huang, Yanping and Cheng, Youlong and Bapna, Ankur and Firat, Orhan and Chen, Dehao and Chen, Mia and Lee, HyoukJoong and Ngiam, Jiquan and Le, Quoc V and Wu, Yonghui and others},
  journal={Advances in neural information processing systems},
  volume={32},
  year={2019}
}

@article{hubert2002linux,
  title={Linux advanced routing \& traffic control HOWTO},
  author={Hubert, Bert and others},
  journal={Netherlabs BV},
  volume={1},
  pages={99--107},
  year={2002}
}

@article{yang2025qwen3,
  title={Qwen3 technical report},
  author={Yang, An and Li, Anfeng and Yang, Baosong and Zhang, Beichen and Hui, Binyuan and Zheng, Bo and Yu, Bowen and Gao, Chang and Huang, Chengen and Lv, Chenxu and others},
  journal={arXiv preprint arXiv:2505.09388},
  year={2025}
}

@article{dubey2024llama,
  title={The llama 3 herd of models},
  author={Dubey, Abhimanyu and Jauhri, Abhinav and Pandey, Abhinav and Kadian, Abhishek and Al-Dahle, Ahmad and Letman, Aiesha and Mathur, Akhil and Schelten, Alan and Yang, Amy and Fan, Angela and others},
  journal={arXiv e-prints},
  pages={arXiv--2407},
  year={2024}
}

@article{merity2016pointer,
  title={Pointer sentinel mixture models},
  author={Merity, Stephen and Xiong, Caiming and Bradbury, James and Socher, Richard},
  journal={arXiv preprint arXiv:1609.07843},
  year={2016}
}

@article{liu2024deepseek,
  title={Deepseek-v3 technical report},
  author={Liu, Aixin and Feng, Bei and Xue, Bing and Wang, Bingxuan and Wu, Bochao and Lu, Chengda and Zhao, Chenggang and Deng, Chengqi and Zhang, Chenyu and Ruan, Chong and others},
  journal={arXiv preprint arXiv:2412.19437},
  year={2024}
}

@article{radford2018improving,
  title={Improving language understanding by generative pre-training},
  author={Radford, Alec and Narasimhan, Karthik and Salimans, Tim and Sutskever, Ilya and others},
  year={2018},
  publisher={San Francisco, CA, USA}
}

@article{king2024sasha,
  title={Sasha: creative goal-oriented reasoning in smart homes with large language models},
  author={King, Evan and Yu, Haoxiang and Lee, Sangsu and Julien, Christine},
  journal={Proceedings of the ACM on Interactive, Mobile, Wearable and Ubiquitous Technologies},
  volume={8},
  number={1},
  pages={1--38},
  year={2024},
  publisher={ACM New York, NY, USA}
}

@article{team2025gemini,
  title={Gemini robotics: Bringing ai into the physical world},
  author={Team, Gemini Robotics and Abeyruwan, Saminda and Ainslie, Joshua and Alayrac, Jean-Baptiste and Arenas, Montserrat Gonzalez and Armstrong, Travis and Balakrishna, Ashwin and Baruch, Robert and Bauza, Maria and Blokzijl, Michiel and others},
  journal={arXiv preprint arXiv:2503.20020},
  year={2025}
}

@article{lin2024awq,
  title={Awq: Activation-aware weight quantization for on-device llm compression and acceleration},
  author={Lin, Ji and Tang, Jiaming and Tang, Haotian and Yang, Shang and Chen, Wei-Ming and Wang, Wei-Chen and Xiao, Guangxuan and Dang, Xingyu and Gan, Chuang and Han, Song},
  journal={Proceedings of machine learning and systems},
  volume={6},
  pages={87--100},
  year={2024}
}

@inproceedings{hsieh2023distilling,
  title={Distilling step-by-step! outperforming larger language models with less training data and smaller model sizes},
  author={Hsieh, Cheng-Yu and Li, Chun-Liang and Yeh, Chih-Kuan and Nakhost, Hootan and Fujii, Yasuhisa and Ratner, Alex and Krishna, Ranjay and Lee, Chen-Yu and Pfister, Tomas},
  booktitle={Findings of the Association for Computational Linguistics: ACL 2023},
  pages={8003--8017},
  year={2023}
}

@article{addo2018credit,
  title={Credit risk analysis using machine and deep learning models},
  author={Addo, Peter Martey and Guegan, Dominique and Hassani, Bertrand},
  journal={Risks},
  volume={6},
  number={2},
  pages={38},
  year={2018},
  publisher={Multidisciplinary Digital Publishing Institute}
}

@article{li2023medical,
  title={Medical image analysis using deep learning algorithms},
  author={Li, Mengfang and Jiang, Yuanyuan and Zhang, Yanzhou and Zhu, Haisheng},
  journal={Frontiers in public health},
  volume={11},
  pages={1273253},
  year={2023},
  publisher={Frontiers Media SA}
}

@inproceedings{hu2022pipeedge,
  title={Pipeedge: Pipeline parallelism for large-scale model inference on heterogeneous edge devices},
  author={Hu, Yang and Imes, Connor and Zhao, Xuanang and Kundu, Souvik and Beerel, Peter A and Crago, Stephen P and Walters, John Paul},
  booktitle={2022 25th Euromicro Conference on Digital System Design (DSD)},
  pages={298--307},
  year={2022},
  organization={IEEE}
}

@inproceedings{sheng2023flexgen,
  title={Flexgen: High-throughput generative inference of large language models with a single gpu},
  author={Sheng, Ying and Zheng, Lianmin and Yuan, Binhang and Li, Zhuohan and Ryabinin, Max and Chen, Beidi and Liang, Percy and R{\'e}, Christopher and Stoica, Ion and Zhang, Ce},
  booktitle={International Conference on Machine Learning},
  pages={31094--31116},
  year={2023},
  organization={PMLR}
}

@inproceedings{aminabadi2022deepspeed,
  title={Deepspeed-inference: enabling efficient inference of transformer models at unprecedented scale},
  author={Aminabadi, Reza Yazdani and Rajbhandari, Samyam and Awan, Ammar Ahmad and Li, Cheng and Li, Du and Zheng, Elton and Ruwase, Olatunji and Smith, Shaden and Zhang, Minjia and Rasley, Jeff and others},
  booktitle={SC22: International Conference for High Performance Computing, Networking, Storage and Analysis},
  pages={1--15},
  year={2022},
  organization={IEEE}
}

@article{gu2023minillm,
  title={Minillm: Knowledge distillation of large language models},
  author={Gu, Yuxian and Dong, Li and Wei, Furu and Huang, Minlie},
  journal={arXiv preprint arXiv:2306.08543},
  year={2023}
}

@inproceedings{liang2023less,
  title={Less is more: Task-aware layer-wise distillation for language model compression},
  author={Liang, Chen and Zuo, Simiao and Zhang, Qingru and He, Pengcheng and Chen, Weizhu and Zhao, Tuo},
  booktitle={International Conference on Machine Learning},
  pages={20852--20867},
  year={2023},
  organization={PMLR}
}

@article{frantar2022gptq,
  title={Gptq: Accurate post-training quantization for generative pre-trained transformers},
  author={Frantar, Elias and Ashkboos, Saleh and Hoefler, Torsten and Alistarh, Dan},
  journal={arXiv preprint arXiv:2210.17323},
  year={2022}
}

@article{ma2023llm,
  title={Llm-pruner: On the structural pruning of large language models},
  author={Ma, Xinyin and Fang, Gongfan and Wang, Xinchao},
  journal={Advances in neural information processing systems},
  volume={36},
  pages={21702--21720},
  year={2023}
}

@inproceedings{frantar2023sparsegpt,
  title={Sparsegpt: Massive language models can be accurately pruned in one-shot},
  author={Frantar, Elias and Alistarh, Dan},
  booktitle={International conference on machine learning},
  pages={10323--10337},
  year={2023},
  organization={PMLR}
}

@article{dettmers2023spqr,
  title={Spqr: A sparse-quantized representation for near-lossless llm weight compression},
  author={Dettmers, Tim and Svirschevski, Ruslan and Egiazarian, Vage and Kuznedelev, Denis and Frantar, Elias and Ashkboos, Saleh and Borzunov, Alexander and Hoefler, Torsten and Alistarh, Dan},
  journal={arXiv preprint arXiv:2306.03078},
  year={2023}
}

@article{hooper2024kvquant,
  title={Kvquant: Towards 10 million context length llm inference with kv cache quantization},
  author={Hooper, Coleman and Kim, Sehoon and Mohammadzadeh, Hiva and Mahoney, Michael W and Shao, Yakun S and Keutzer, Kurt and Gholami, Amir},
  journal={Advances in Neural Information Processing Systems},
  volume={37},
  pages={1270--1303},
  year={2024}
}

@article{dettmers2023qlora,
  title={Qlora: Efficient finetuning of quantized llms},
  author={Dettmers, Tim and Pagnoni, Artidoro and Holtzman, Ari and Zettlemoyer, Luke},
  journal={Advances in neural information processing systems},
  volume={36},
  pages={10088--10115},
  year={2023}
}

@article{ashkboos2024slicegpt,
  title={Slicegpt: Compress large language models by deleting rows and columns},
  author={Ashkboos, Saleh and Croci, Maximilian L and Nascimento, Marcelo Gennari do and Hoefler, Torsten and Hensman, James},
  journal={arXiv preprint arXiv:2401.15024},
  year={2024}
}

@article{li2025unity,
  title={Unity is power: Semi-asynchronous collaborative training of large-scale models with structured pruning in resource-limited clients},
  author={Li, Yan and Zhang, Xiao and Li, Mingyi and Xu, Guangwei and Chen, Feng and Yuan, Yuan and Zou, Yifei and Zhao, Mengying and Lu, Jianbo and Yu, Dongxiao},
  journal={IEEE Transactions on Mobile Computing},
  year={2025},
  publisher={IEEE}
}

@article{zeng2020coedge,
  title={Coedge: Cooperative dnn inference with adaptive workload partitioning over heterogeneous edge devices},
  author={Zeng, Liekang and Chen, Xu and Zhou, Zhi and Yang, Lei and Zhang, Junshan},
  journal={IEEE/ACM Transactions on Networking},
  volume={29},
  number={2},
  pages={595--608},
  year={2020},
  publisher={IEEE}
}

@inproceedings{jeong2022band,
  title={Band: coordinated multi-dnn inference on heterogeneous mobile processors},
  author={Jeong, Joo Seong and Lee, Jingyu and Kim, Donghyun and Jeon, Changmin and Jeong, Changjin and Lee, Youngki and Chun, Byung-Gon},
  booktitle={Proceedings of the 20th Annual International Conference on Mobile Systems, Applications and Services},
  pages={235--247},
  year={2022}
}

@inproceedings{hu2024edge,
  title={When the edge meets transformers: Distributed inference with transformer models},
  author={Hu, Chenghao and Li, Baochun},
  booktitle={2024 IEEE 44th International Conference on Distributed Computing Systems (ICDCS)},
  pages={82--92},
  year={2024},
  organization={IEEE}
}

@article{li2024resource,
  title={Resource-aware federated self-supervised learning with global class representations},
  author={Li, Mingyi and Zhang, Xiao and Wang, Qi and Liu, Tengfei and Wu, Ruofan and Wang, Weiqiang and Zhuang, Fuzhen and Xiong, Hui and Yu, Dongxiao},
  journal={Advances in Neural Information Processing Systems},
  volume={37},
  pages={10008--10035},
  year={2024}
}

@inproceedings{wang2023theoretical,
  title={Theoretical convergence guaranteed resource-adaptive federated learning with mixed heterogeneity},
  author={Wang, Yangyang and Zhang, Xiao and Li, Mingyi and Lan, Tian and Chen, Huashan and Xiong, Hui and Cheng, Xiuzhen and Yu, Dongxiao},
  booktitle={Proceedings of the 29th ACM SIGKDD Conference on Knowledge Discovery and Data Mining},
  pages={2444--2455},
  year={2023}
}

@inproceedings{li2025generalization,
  title={Generalization-Aware Distributed Minimax Optimization for Large-Scale Models on Resource-Limited Devices},
  author={Li, Mingyi and Zhang, Xiao and Yuan, Yuan and Zou, Yifei and Guo, Shaoyong and Cheng, Xiuzhen and Yu, Dongxiao},
  booktitle={The 21st International Conference on Mobility, Sensing and Networking},
  year={2025}
}

@misc{cuda_event,
  author       = {{NVIDIA}},
  title        = {{CUDA Runtime API}},
  howpublished = {[Online]. Available: \url{https://docs.nvidia.com/cuda/archive/12.2.2/pdf/CUDA_Runtime_API.pdf}},
  year         = {2023},
  note         = {Version 12.2.2, accessed: Apr. 23, 2026}
}

@misc{io_uring,
  author       = {{Linux man-pages Project}},
  title        = {{io\_uring(7): Linux Manual Page}},
  howpublished = {[Online]. Available: \url{https://man7.org/linux/man-pages/man7/io_uring.7.html}},
  year         = {2025},
  note         = {Accessed: Aug. 30, 2026}
}

@misc{nvidia_jetson_xavier_nx,
  author       = {{NVIDIA}},
  title        = {{Introducing Jetson Xavier NX, the World's Smallest AI Supercomputer}},
  howpublished = {\url{https://developer.nvidia.com/blog/jetson-xavier-nx-the-worlds-smallest-ai-supercomputer/}},
  year         = {2019},
  note         = {Accessed: 2026-06-07}
}

@misc{hf_accelerate_big_modeling,
  author       = {{Hugging Face}},
  title        = {{Accelerate: Big Model Inference}},
  howpublished = {\url{https://huggingface.co/docs/accelerate/usage_guides/big_modeling}},
  note         = {Accessed: 2026-06-08}
}
\begin{IEEEbiography}[{\includegraphics[width=1in,height=1.25in,clip,keepaspectratio]{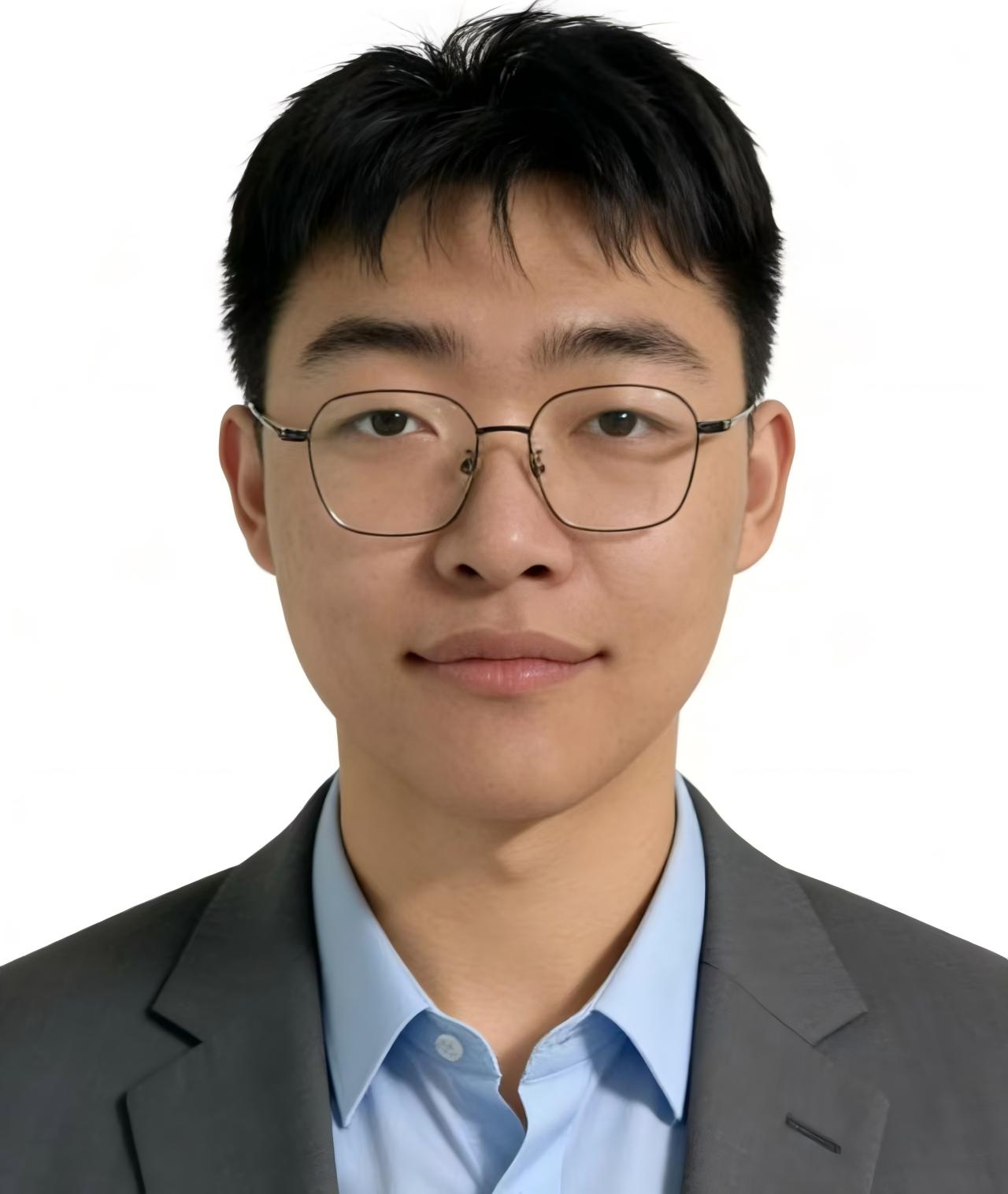}}]{Mingyu Sun} is currently pursuing the Ph.D. degree with the School of Computer Science and Technology, Shandong University. He received his B.S. degree in Shandong University. His research interests include collaborative inference and distributed systems.
\end{IEEEbiography}
\vspace{-20pt}
\begin{IEEEbiography}[{\includegraphics[width=1in,height=1.25in,clip,keepaspectratio]{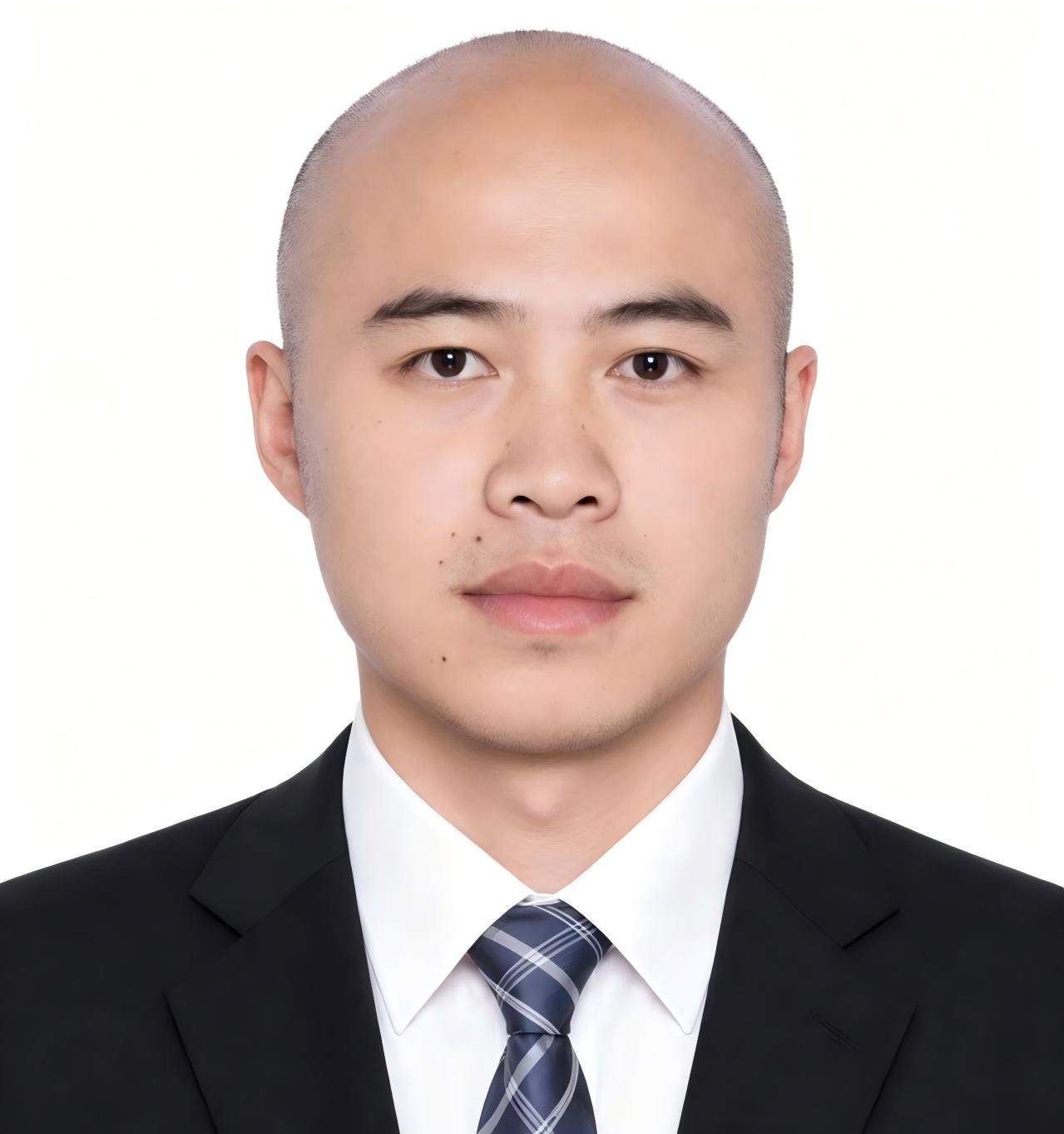}}]{Xiao Zhang} is now an associate professor in the School of Computer Science and Technology, Shandong University. His research interests include distributed learning and federated learning. 
He has published more than 70 papers in the prestigious refereed journals and conference proceedings, such as IEEE Transactions on Knowledge and Data Engineering, IEEE Transactions on Mobile Computing, ACM SIGKDD, ICML, NeurIPS, and UBICOMP.  
\end{IEEEbiography}
\vspace{-20pt}
\begin{IEEEbiography}[{\includegraphics[width=1in,height=1.25in,clip,keepaspectratio]{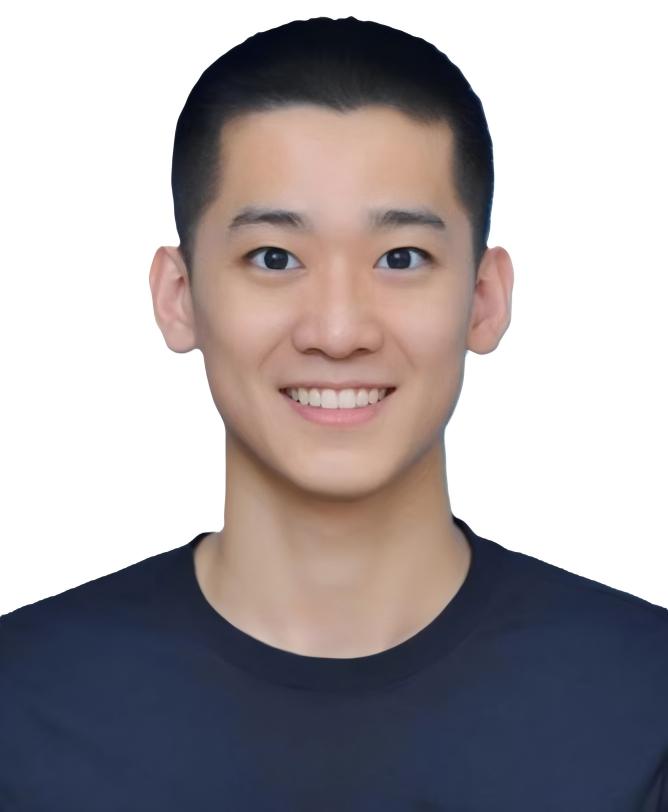}}]{Shen Qu} is currently an undergraduate student in the School of Computer Science and Technology, Shandong University. His research interests include distributed learning and edge computing.
\end{IEEEbiography}
\vspace{-20pt}
\begin{IEEEbiography}[{\includegraphics[width=1in,height=1.25in,clip,keepaspectratio]{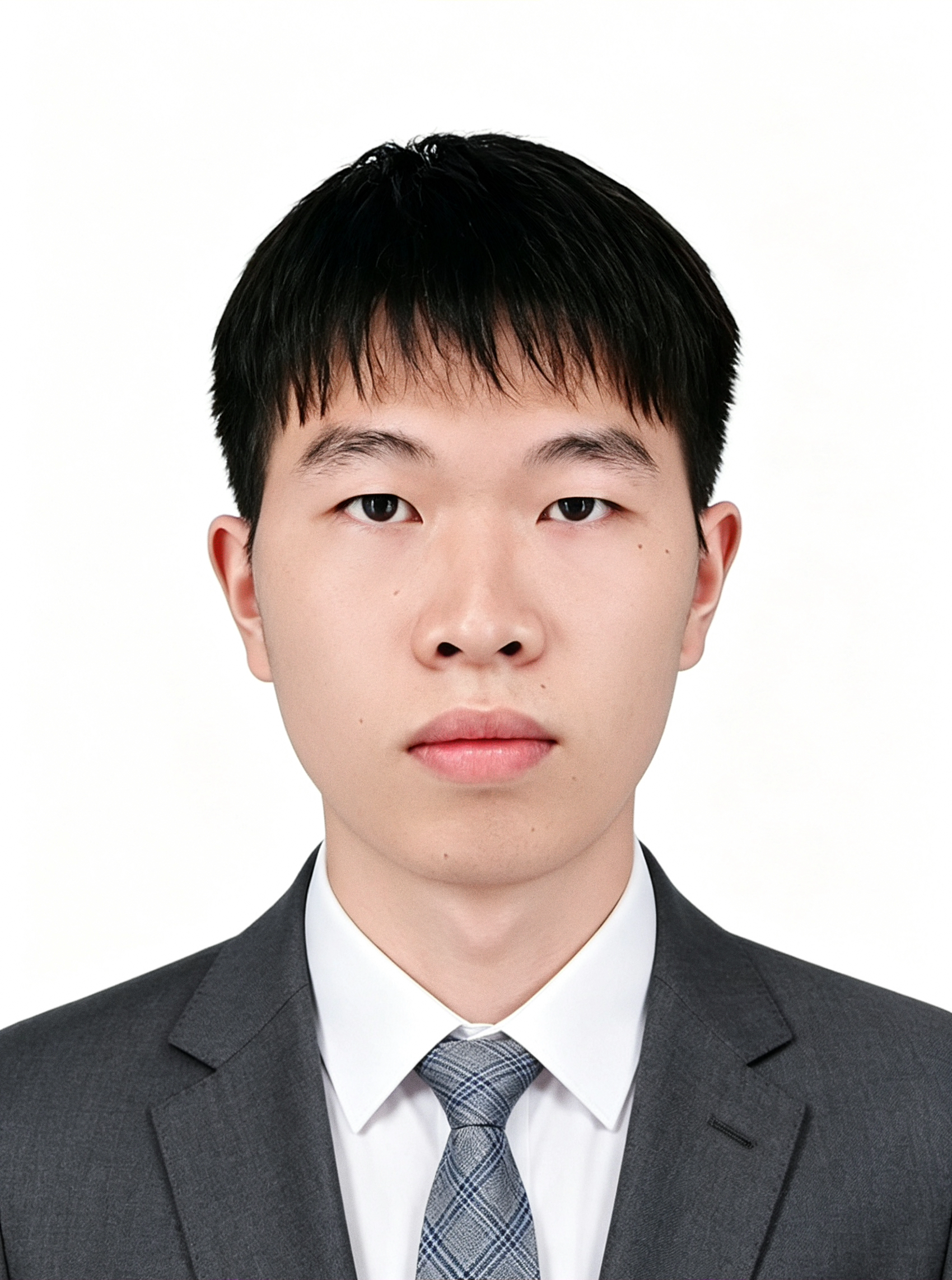}}]{Yan Li} is currently pursuing the Ph.D. degree with the School of Computer Science and Technology, Shandong University. He received his B.S. degree in Shandong University. His research interests include collaborative learning and distributed systems.
\end{IEEEbiography}
\vspace{-20pt}
\begin{IEEEbiography}[{\includegraphics[width=1in,height=1.25in,clip,keepaspectratio]{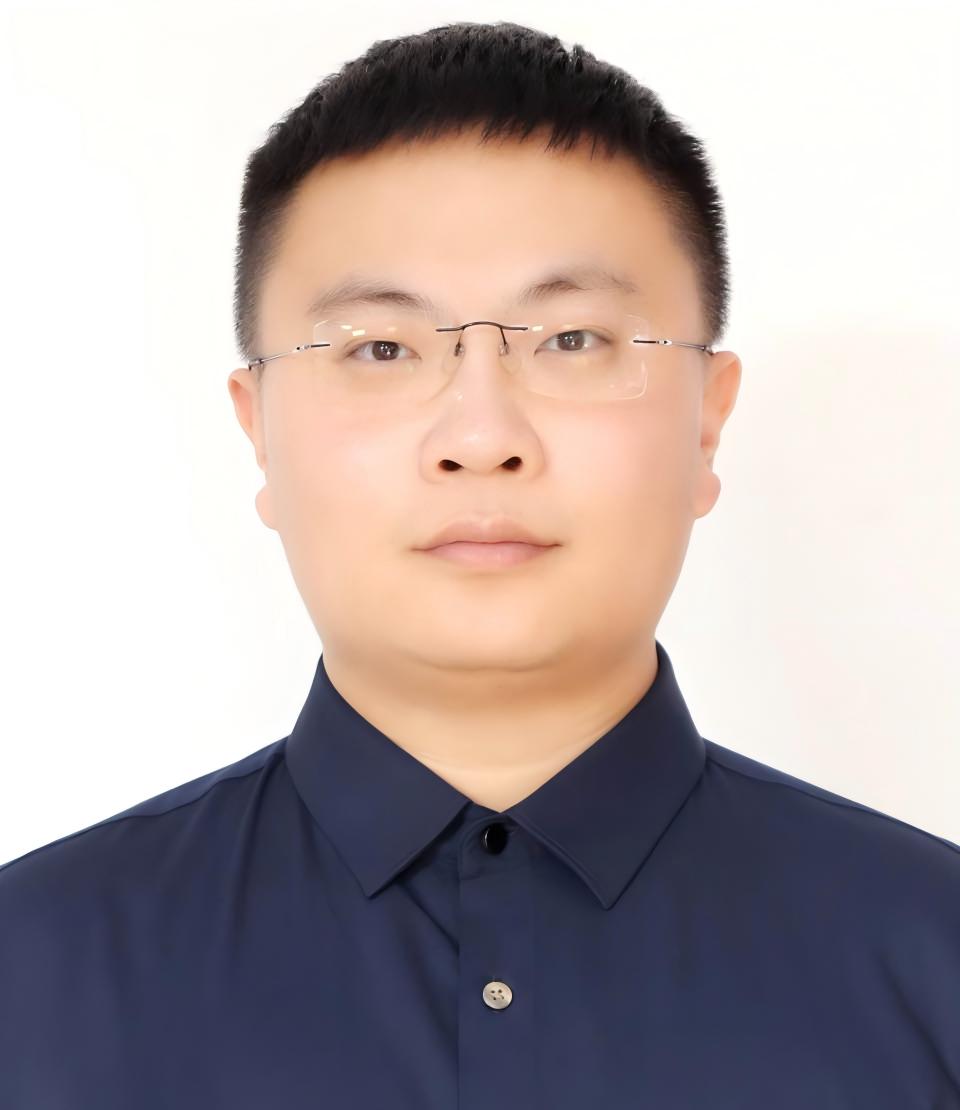}}]{Mengbai Xiao}, Ph.D., is a Professor in the School of Computer Science and Technology at Shandong University, China. He received the Ph.D. degree in Computer Science from George Mason University in 2018, and the M.S. degree in Software Engineering from University of Science and Technology of China in 2011. He was a postdoctoral researcher at the HPCS Lab, the Ohio State University. His research interests include multimedia systems, parallel and distributed systems. He has published papers in prestigious conferences such as USENIX ATC, ACM Multimedia, IEEE ICDE, IEEE ICDCS, IEEE INFOCOM.
\end{IEEEbiography}
\vspace{-20pt}
\begin{IEEEbiography}[{\includegraphics[width=1in,height=1.25in,clip,keepaspectratio]{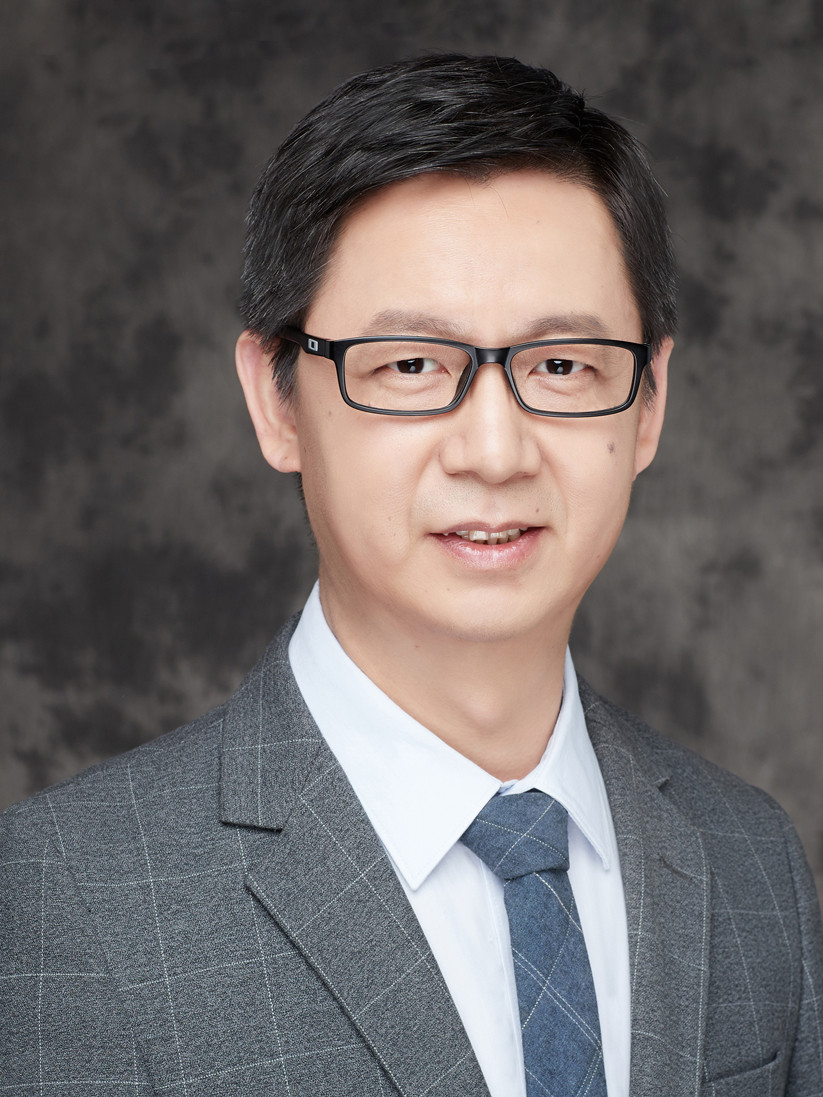}}]{Yanwei Zheng} received the Ph.D. degree from Beihang University in January 2019, supervised by Prof. Zhang Xiong. He is currently an Associate Professor in the School of Computer Science and Technology, Shandong University. His research interests include computer vision, visual-and-language navigation, and object-goal visual navigation.
\end{IEEEbiography}
\vspace{-20pt}
\begin{IEEEbiography}[{\includegraphics[width=1in,height=1.25in,clip,keepaspectratio]{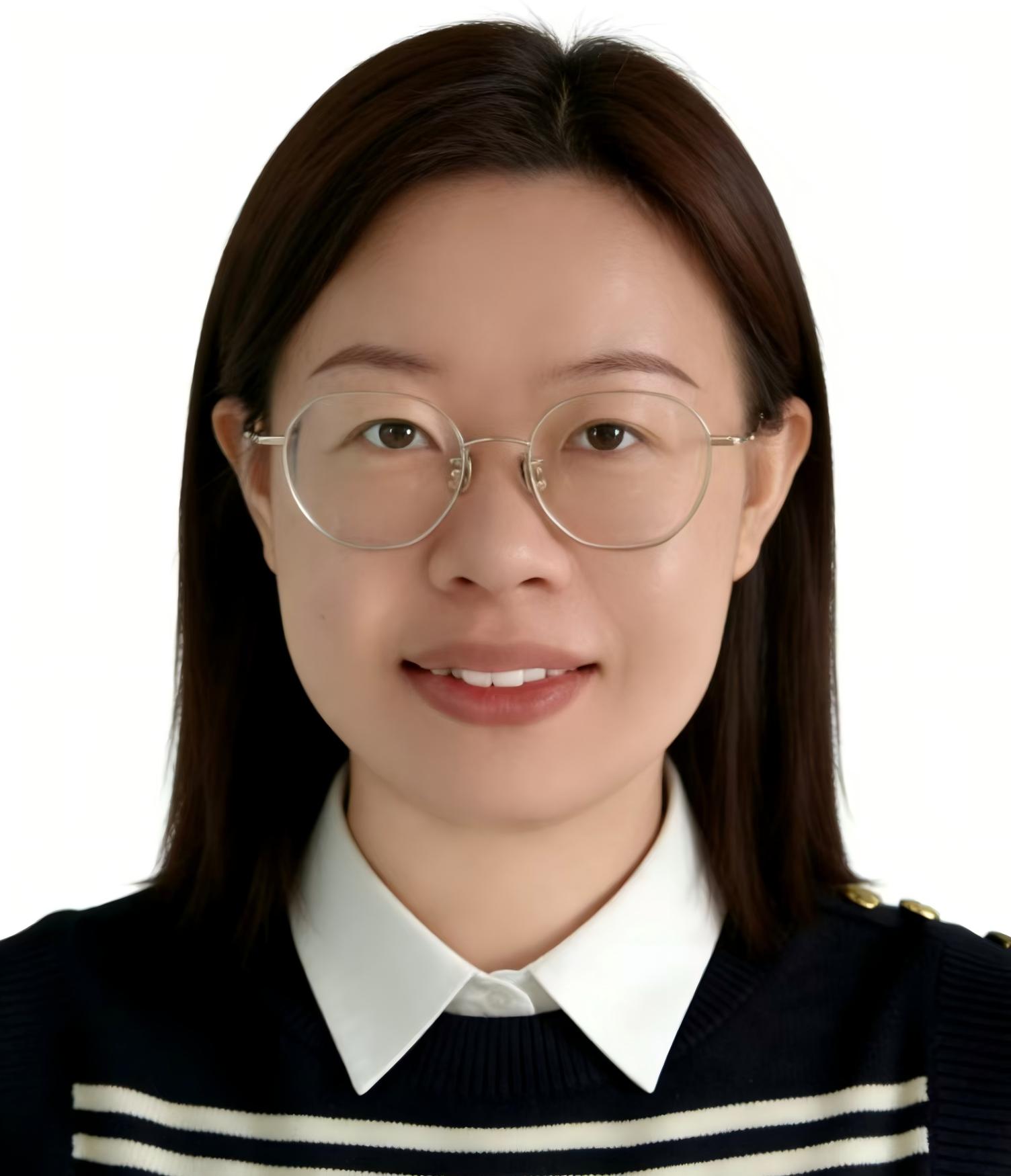}}]{Yuan Yuan} received the BSc degrees from the School of Mathematical Sciences, Shanxi University in 2016, and the Ph.D. degree from the School of  Computer Science and Technology, Shandong University, Qingdao, China, in 2021. She is currently an Associate Professor with School of Artificial Intelligence, Shandong University. Her research interests include distributed computing and distributed machine learning.
\end{IEEEbiography}
\vspace{-20pt}
\begin{IEEEbiography}[{\includegraphics[width=1in,height=1.25in,clip,keepaspectratio]{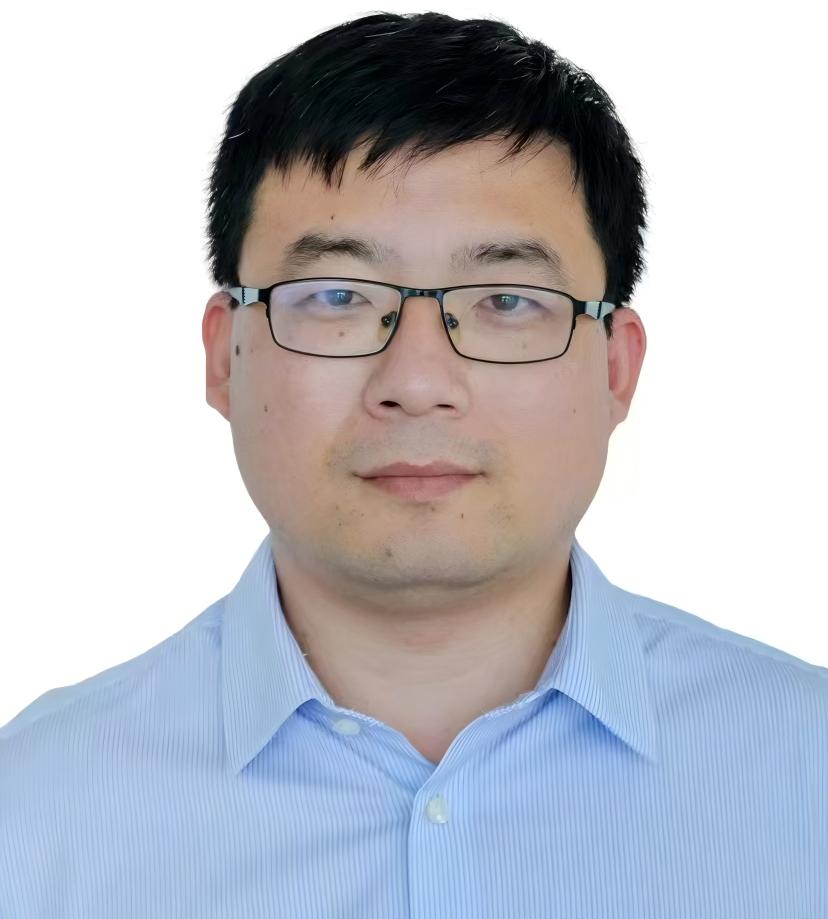}}]{Dongxiao Yu} received the B.S. degree in 2006 from the School of Mathematics, Shandong University and the Ph.D degree in 2014 from the Department of Computer Science, The University of Hong Kong. He became an associate professor in the School of Computer Science and Technology, Huazhong University of Science and Technology, in 2016. He is currently a professor in the School of Cryptologic Science and Engineering, Shandong University. His research interests include edge intelligence, distributed computing and data mining.
\end{IEEEbiography}

\vfill

\end{document}